\documentclass[twocolumn,letterpaper,aps,prc,longbibliography,superscriptaddress,nofootinbib,floatfix]{revtex4-1}

\usepackage{graphicx}	
\usepackage{xspace}     

\newcommand{\sqs}{\mbox{$\sqrt{s}$}\xspace}
\newcommand{\sqsn}{\mbox{$\sqrt{s_{_{NN}}}$}\xspace}

\newcommand{\pp}{\mbox{$p$$+$$p$}\xspace}

\newcommand{\pau}{\mbox{$p$$+$Au}\xspace}
\newcommand{\pal}{\mbox{$p$$+$Al}\xspace}
\newcommand{\heau}{\mbox{$^{3}$He$+$Au}\xspace}

\begin{document}

\title{Measurement of the relative yields of $\psi(2S)$ to $\psi(1S)$ 
mesons produced at forward and backward rapidity in 
$p$$+$$p$, $p$$+$Al, $p$$+$Au, and $^{3}$He$+$Au collisions at 
$\sqrt{s_{_{NN}}}=200$ GeV}

\newcommand{\abilene}{Abilene Christian University, Abilene, Texas 79699, USA}
\newcommand{\augie}{Department of Physics, Augustana University, Sioux Falls, South Dakota 57197, USA}
\newcommand{\banaras}{Department of Physics, Banaras Hindu University, Varanasi 221005, India}
\newcommand{\barc}{Bhabha Atomic Research Centre, Bombay 400 085, India}
\newcommand{\baruch}{Baruch College, City University of New York, New York, New York, 10010 USA}
\newcommand{\bnlcoll}{Collider-Accelerator Department, Brookhaven National Laboratory, Upton, New York 11973-5000, USA}
\newcommand{\bnlphys}{Physics Department, Brookhaven National Laboratory, Upton, New York 11973-5000, USA}
\newcommand{\caucr}{University of California-Riverside, Riverside, California 92521, USA}
\newcommand{\charlesczech}{Charles University, Ovocn\'{y} trh 5, Praha 1, 116 36, Prague, Czech Republic}
\newcommand{\chonbuk}{Chonbuk National University, Jeonju, 561-756, Korea}
\newcommand{\ciae}{Science and Technology on Nuclear Data Laboratory, China Institute of Atomic Energy, Beijing 102413, People's Republic of China}
\newcommand{\cns}{Center for Nuclear Study, Graduate School of Science, University of Tokyo, 7-3-1 Hongo, Bunkyo, Tokyo 113-0033, Japan}
\newcommand{\colorado}{University of Colorado, Boulder, Colorado 80309, USA}
\newcommand{\columbia}{Columbia University, New York, New York 10027 and Nevis Laboratories, Irvington, New York 10533, USA}
\newcommand{\czechtech}{Czech Technical University, Zikova 4, 166 36 Prague 6, Czech Republic}
\newcommand{\debrecen}{Debrecen University, H-4010 Debrecen, Egyetem t{\'e}r 1, Hungary}
\newcommand{\elte}{ELTE, E{\"o}tv{\"o}s Lor{\'a}nd University, H-1117 Budapest, P{\'a}zm{\'a}ny P.~s.~1/A, Hungary}
\newcommand{\eszterhazy}{Eszterh\'azy K\'aroly University, K\'aroly R\'obert Campus, H-3200 Gy\"ngy\"os, M\'atrai \'ut 36, Hungary}
\newcommand{\ewha}{Ewha Womans University, Seoul 120-750, Korea}
\newcommand{\fsu}{Florida State University, Tallahassee, Florida 32306, USA}
\newcommand{\gsu}{Georgia State University, Atlanta, Georgia 30303, USA}
\newcommand{\hiroshima}{Hiroshima University, Kagamiyama, Higashi-Hiroshima 739-8526, Japan}
\newcommand{\howard}{Department of Physics and Astronomy, Howard University, Washington, DC 20059, USA}
\newcommand{\ihepprot}{IHEP Protvino, State Research Center of Russian Federation, Institute for High Energy Physics, Protvino, 142281, Russia}
\newcommand{\illuiuc}{University of Illinois at Urbana-Champaign, Urbana, Illinois 61801, USA}
\newcommand{\inrras}{Institute for Nuclear Research of the Russian Academy of Sciences, prospekt 60-letiya Oktyabrya 7a, Moscow 117312, Russia}
\newcommand{\instpasczech}{Institute of Physics, Academy of Sciences of the Czech Republic, Na Slovance 2, 182 21 Prague 8, Czech Republic}
\newcommand{\isu}{Iowa State University, Ames, Iowa 50011, USA}
\newcommand{\jaea}{Advanced Science Research Center, Japan Atomic Energy Agency, 2-4 Shirakata Shirane, Tokai-mura, Naka-gun, Ibaraki-ken 319-1195, Japan}
\newcommand{\jyvaskyla}{Helsinki Institute of Physics and University of Jyv{\"a}skyl{\"a}, P.O.Box 35, FI-40014 Jyv{\"a}skyl{\"a}, Finland}
\newcommand{\kek}{KEK, High Energy Accelerator Research Organization, Tsukuba, Ibaraki 305-0801, Japan}
\newcommand{\korea}{Korea University, Seoul, 136-701, Korea}
\newcommand{\kurchatov}{National Research Center ``Kurchatov Institute", Moscow, 123098 Russia}
\newcommand{\kyoto}{Kyoto University, Kyoto 606-8502, Japan}
\newcommand{\lawllnl}{Lawrence Livermore National Laboratory, Livermore, California 94550, USA}
\newcommand{\losalamos}{Los Alamos National Laboratory, Los Alamos, New Mexico 87545, USA}
\newcommand{\lund}{Department of Physics, Lund University, Box 118, SE-221 00 Lund, Sweden}
\newcommand{\maryland}{University of Maryland, College Park, Maryland 20742, USA}
\newcommand{\mass}{Department of Physics, University of Massachusetts, Amherst, Massachusetts 01003-9337, USA}
\newcommand{\michigan}{Department of Physics, University of Michigan, Ann Arbor, Michigan 48109-1040, USA}
\newcommand{\muhlenberg}{Muhlenberg College, Allentown, Pennsylvania 18104-5586, USA}
\newcommand{\nara}{Nara Women's University, Kita-uoya Nishi-machi Nara 630-8506, Japan}
\newcommand{\natmephi}{National Research Nuclear University, MEPhI, Moscow Engineering Physics Institute, Moscow, 115409, Russia}
\newcommand{\newmex}{University of New Mexico, Albuquerque, New Mexico 87131, USA}
\newcommand{\nmsu}{New Mexico State University, Las Cruces, New Mexico 88003, USA}
\newcommand{\ohio}{Department of Physics and Astronomy, Ohio University, Athens, Ohio 45701, USA}
\newcommand{\ornl}{Oak Ridge National Laboratory, Oak Ridge, Tennessee 37831, USA}
\newcommand{\orsay}{IPN-Orsay, Univ.~Paris-Sud, CNRS/IN2P3, Universit\'e Paris-Saclay, BP1, F-91406, Orsay, France}
\newcommand{\peking}{Peking University, Beijing 100871, People's Republic of China}
\newcommand{\pnpi}{PNPI, Petersburg Nuclear Physics Institute, Gatchina, Leningrad region, 188300, Russia}
\newcommand{\riken}{RIKEN Nishina Center for Accelerator-Based Science, Wako, Saitama 351-0198, Japan}
\newcommand{\rikjrbrc}{RIKEN BNL Research Center, Brookhaven National Laboratory, Upton, New York 11973-5000, USA}
\newcommand{\rikkyo}{Physics Department, Rikkyo University, 3-34-1 Nishi-Ikebukuro, Toshima, Tokyo 171-8501, Japan}
\newcommand{\saispbstu}{Saint Petersburg State Polytechnic University, St.~Petersburg, 195251 Russia}
\newcommand{\seoulnat}{Department of Physics and Astronomy, Seoul National University, Seoul 151-742, Korea}
\newcommand{\stonybrkc}{Chemistry Department, Stony Brook University, SUNY, Stony Brook, New York 11794-3400, USA}
\newcommand{\stonycrkp}{Department of Physics and Astronomy, Stony Brook University, SUNY, Stony Brook, New York 11794-3800, USA}
\newcommand{\tenn}{University of Tennessee, Knoxville, Tennessee 37996, USA}
\newcommand{\titech}{Department of Physics, Tokyo Institute of Technology, Oh-okayama, Meguro, Tokyo 152-8551, Japan}
\newcommand{\tsukuba}{Center for Integrated Research in Fundamental Science and Engineering, University of Tsukuba, Tsukuba, Ibaraki 305, Japan}
\newcommand{\vandy}{Vanderbilt University, Nashville, Tennessee 37235, USA}
\newcommand{\weizmann}{Weizmann Institute, Rehovot 76100, Israel}
\newcommand{\wigner}{Institute for Particle and Nuclear Physics, Wigner Research Centre for Physics, Hungarian Academy of Sciences (Wigner RCP, RMKI) H-1525 Budapest 114, POBox 49, Budapest, Hungary}
\newcommand{\yonsei}{Yonsei University, IPAP, Seoul 120-749, Korea}
\newcommand{\zagreb}{University of Zagreb, Faculty of Science, Department of Physics, Bijeni\v{c}ka 32, HR-10002 Zagreb, Croatia}
\affiliation{\abilene}
\affiliation{\augie}
\affiliation{\banaras}
\affiliation{\barc}
\affiliation{\baruch}
\affiliation{\bnlcoll}
\affiliation{\bnlphys}
\affiliation{\caucr}
\affiliation{\charlesczech}
\affiliation{\chonbuk}
\affiliation{\ciae}
\affiliation{\cns}
\affiliation{\colorado}
\affiliation{\columbia}
\affiliation{\czechtech}
\affiliation{\debrecen}
\affiliation{\elte}
\affiliation{\eszterhazy}
\affiliation{\ewha}
\affiliation{\fsu}
\affiliation{\gsu}
\affiliation{\hiroshima}
\affiliation{\howard}
\affiliation{\ihepprot}
\affiliation{\illuiuc}
\affiliation{\inrras}
\affiliation{\instpasczech}
\affiliation{\isu}
\affiliation{\jaea}
\affiliation{\jyvaskyla}
\affiliation{\kek}
\affiliation{\korea}
\affiliation{\kurchatov}
\affiliation{\kyoto}
\affiliation{\lawllnl}
\affiliation{\losalamos}
\affiliation{\lund}
\affiliation{\maryland}
\affiliation{\mass}
\affiliation{\michigan}
\affiliation{\muhlenberg}
\affiliation{\nara}
\affiliation{\natmephi}
\affiliation{\newmex}
\affiliation{\nmsu}
\affiliation{\ohio}
\affiliation{\ornl}
\affiliation{\orsay}
\affiliation{\peking}
\affiliation{\pnpi}
\affiliation{\riken}
\affiliation{\rikjrbrc}
\affiliation{\rikkyo}
\affiliation{\saispbstu}
\affiliation{\seoulnat}
\affiliation{\stonybrkc}
\affiliation{\stonycrkp}
\affiliation{\tenn}
\affiliation{\titech}
\affiliation{\tsukuba}
\affiliation{\vandy}
\affiliation{\weizmann}
\affiliation{\wigner}
\affiliation{\yonsei}
\affiliation{\zagreb}
\author{A.~Adare} \affiliation{\colorado} 
\author{C.~Aidala} \affiliation{\michigan} 
\author{N.N.~Ajitanand} \affiliation{\stonybrkc} 
\author{Y.~Akiba} \email[PHENIX Spokesperson: ]{akiba@rcf.rhic.bnl.gov} \affiliation{\riken} \affiliation{\rikjrbrc} 
\author{M.~Alfred} \affiliation{\howard} 
\author{V.~Andrieux} \affiliation{\michigan} 
\author{K.~Aoki} \affiliation{\kek} \affiliation{\riken} 
\author{N.~Apadula} \affiliation{\isu} \affiliation{\stonycrkp} 
\author{H.~Asano} \affiliation{\kyoto} \affiliation{\riken} 
\author{C.~Ayuso} \affiliation{\michigan} 
% ...
\collaboration{PHENIX Collaboration} \noaffiliation

\begin{abstract}

The PHENIX Collaboration has measured the ratio of the yields of 
$\psi(2S)$ to $\psi(1S)$ mesons produced in $p$$+$$p$, $p$$+$Al, $p$$+$Au, 
and $^{3}$He$+$Au collisions at $\sqrt{s_{_{NN}}}=200$ GeV over the 
forward and backward rapidity intervals $1.2<|y|<2.2$.  We find that the 
ratio in $p$$+$$p$ collisions is consistent with measurements at other 
collision energies.  In collisions with nuclei, we find that in the 
forward ($p$-going or $^{3}$He-going) direction, the relative yield of 
$\psi(2S)$ mesons to $\psi(1S)$ mesons is consistent with the value 
measured in \pp collisions.  However, in the backward (nucleus-going) 
direction, the $\psi(2S)$ is preferentially suppressed by a factor of 
$\sim$2.  This suppression is attributed in some models to breakup of the 
weakly-bound $\psi(2S)$ through final state interactions with comoving 
particles, which have a higher density in the nucleus-going direction.  
These breakup effects may compete with color screening in a deconfined 
quark-gluon plasma to produce sequential suppression of excited quarkonia 
states.

\end{abstract}

	
\maketitle


\section{Introduction}

The production of quark-antiquark bound states in nuclear collisions has 
long been studied for evidence of a phase transition between normal 
nuclear matter, where quarks and gluons are confined in hadrons, to a 
plasma phase where colored partons are deconfined.  Early predictions of 
charmonium suppression as an unambiguous signature of 
deconfinement~\cite{Matsui:1986dk} have proven to be overly simplistic, as 
a variety of competing mechanisms have been identified which do not 
require color screening to disrupt bound state $q\bar{q}$ pair formation 
and hadronization.  It is necessary to quantitatively account for these 
effects to correctly interpret what measurements of quarkonia 
suppression in nuclear collisions imply about the quantum-chromodynamics 
phase diagram.
	
The heavy charm and bottom quarks are of particular interest, as they are 
produced through hard processes that are calculable with perturbative 
quantum-chromodynamics techniques~\cite{Cacciari:2005rk} and their bound 
states are accessible experimentally through decays to dileptons.  Models 
of $c\bar{c}$ and $b\bar{b}$ bound state production generally factorize 
quarkonia production into two stages: first, the prompt initial heavy 
quark production via gluon fusion, and after a formation time $\tau_{f}$, 
the mechanism leading to hadronization into the final color singlet 
state~\cite{Fritzsch:1977ay,Bodwin:2005hm, Kang:2014tta}.  In collisions 
involving nuclei, the initial heavy quark production can be affected by 
modifications of the parton distribution functions~\cite{Helenius:2012wd}, 
energy loss in the nucleus~\cite{Vitev:2007ve}, and scattering with other 
partons~\cite{Vogt:2001nh}.  Effects which may be of hydrodynamic origin 
are also present in small 
systems~\cite{Adare:2013piz,Adare:2014keg,Adare:2015ctn}, and may further 
alter the heavy quark final state~\cite{Du:2015wha,Beraudo:2015wsd}.  If 
these flow effects are due to quark-gluon-plasma formation, the presence 
of deconfined colored partons can inhibit coalescence into a bound state 
or dissolve fully-formed bound states~\cite{Mocsy:2013syh}.  The 
fully-formed pair may also be broken up through interactions with 
comoving hadrons outside the nucleus~\cite{Capella:1996va,Ferreiro:2014bia}.
	
One way to isolate final-state effects is through studies of states with 
the same quark content but different binding energies, such as the 
charmonium states $\psi(1S)$ and $\psi(2S)$, with binding energies of 
$\sim$640 and $\sim$50 MeV, respectively~\cite{Satz:2005hx}.  
Before the charmonium formation time $\tau_{f} \sim$0.15 fm/$c$, the 
precursor state is thought to be the same and so any effects on the 
precursor are likely identical.  While significant initial-state effects 
on open charm have been found at the Relativistic Heavy Ion Collider 
(RHIC)~\cite{Adare:2012yxa,Adare:2013lkk}, these should 
equally affect all charm pairs before projection onto a final state.  
Therefore any differences in the modification of $\psi(2S)$ and $\psi(1S)$ 
production are likely due to late time effects which are sensitive to 
differences in the fully-formed meson radius and binding energy.
         
Previous measurements of $\psi(2S)$ suppression in $p$$+A$ collisions by 
the E866/NuSea~\cite{Leitch:1999ea} and NA50~\cite{Alessandro:2006jt} 
experiments were found to be well explained by models based on the breakup 
of fully formed charmonium states inside the nucleus, which naturally 
leads to a larger effect on the $\psi(2S)$ due to its larger 
radius~\cite{Arleo:1999af}.  However, this model fails to reproduce data 
from $d+$Au collisions at midrapidity at RHIC~\cite{Adare:2013ezl}, where 
the higher beam energy and shorter nuclear crossing time means the 
$c\bar{c}$ pairs project onto their final states outside the 
nucleus~\cite{McGlinchey:2012bp}.  Breakup mechanisms which occur after 
the formation time can however explain the different suppression.  
Measurements over different rapidity intervals in asymmetric collisions 
can simultaneously provide information on the evolution of the $c\bar{c}$ 
state in different hadronic environments.  In the $p/d/^{3}$He-going 
direction, there are relatively few produced particles, while in the 
$A$-going direction, there may be significant final state interactions 
between the quarkonia state and the higher number of comoving hadrons.

Here, we present measurements of the relative yields of $\psi(2S)$ 
to $\psi(1S)$ mesons at forward and backward rapidity in \pp, \pal, \pau, 
and \heau collisions at \sqsn = 200 GeV, from the 2014 and 2015 PHENIX data sets.  We find that the relative 
production rate of $\psi(2S)$ to $\psi(1S)$ mesons in \pp collisions is 
consistent with expectations from a modified color evaporation model of 
charmonium production~\cite{Ramona_privcomm}. In $p/^{3}$He$+$$A$ 
collisions, over the forward ($p/^{3}$He-going) rapidity interval 
$1.2<y<2.2$ the relative production rates of the two states are consistent 
with what is found in \pp.  However, over the backward ($A$-going) 
rapidity interval $-2.2<y<-1.2$, the $\psi(2S)$ is preferentially more 
suppressed than the $\psi(1S)$ by a factor of $\sim$2, which is likely due 
to final state effects such as interactions with comoving hadrons.

\section{Experimental Setup}

Dimuons from $\psi(2S)$ and $\psi(1S)$ decays were measured with the 
PHENIX muon spectrometer, which comprises two arms covering the forward 
and backward rapidity intervals $1.2<|y|<2.2$~\cite{Akikawa:2003zs}.  
Muons considered in this analysis pass through a the forward vertex 
tracking detector (FVTX)~\cite{Aidala:2013vna}, a hadron absorber with a 
thickness of $\sim10$ interaction lengths, then through three 
multi-plane cathode strip tracking chambers which reside in a radial 
magnetic field.  After the tracking chambers are more layers of 
absorbers and Iarocci tubes for muon identification, where tracks must 
penetrate at least 2.5 interaction lengths of additional steel absorber. 
The dimuon trigger used in this analysis records events which have two 
tracks in the same spectrometer arm that pass through all absorber 
material.

Previous PHENIX measurements were not able to resolve the smaller 
$\psi(2S)$ peak in the dimuon mass spectrum near the larger, broad 
$\psi(1S)$ peak (cf.~\cite{Adare:2012qf,Adare:2011vq}).  However, the 
introduction of the four silicon tracking layers of the FVTX in 2012 now 
allows a precise measurement of the pair opening angle to be made before 
the muons undergo multiple scattering in the absorber.  The additional 
FVTX tracking improves the dimuon mass resolution and reduces 
combinatorial background from hadron decays, allowing separation of the 
two peaks.

\section{Data Analysis}

\begin{figure}[htbp]
  \centering
  \includegraphics[width=1.0\linewidth]{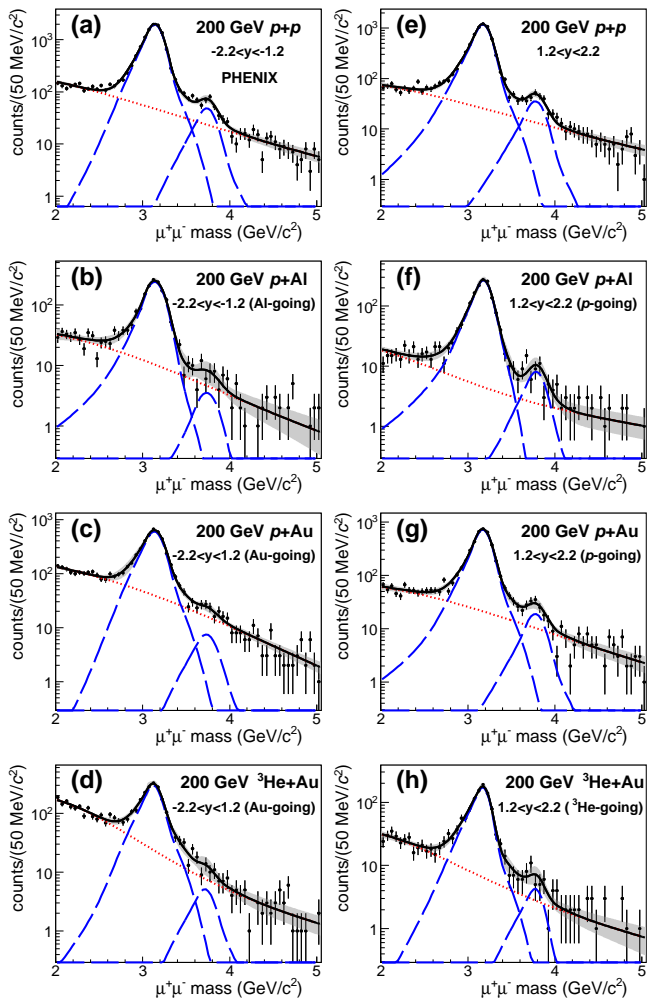}
  \caption{The measured dimuon mass spectra with fits from the (a--d) 
South and (e--h) North PHENIX muon arms, for \pp, \pal, \pau, and \heau 
collisions at \sqsn=200 GeV. The total fit is the solid [black] line with 
a shaded [gray] band representing the 90\% confidence level.  The dashed 
[blue] and dotted [red] lines represent the contributions from the 
resonances and background, respectively.}
  \label{fig:fits}
\end{figure}

The measured dimuon mass spectra from \pp, \pal, \pau, and \heau 
collisions at \sqsn = 200 GeV are shown in Fig.~\ref{fig:fits}, with the 
a--d (e--h) plots showing data recorded in the South (North) PHENIX muon 
spectrometer arm.  These distributions are comprised of peaks at the 
$\psi(1S)$ and $\psi(2S)$ masses ($\sim3.1$ and $\sim3.7$ GeV/$c^{2}$, 
respectively) on top of correlated background from charm and bottom 
hadron decays and Drell-Yan pairs, plus combinatorial background from 
light meson decays and hadrons which are not stopped in the absorbers.  
The fits to the data are shown as a solid [black] line with a shaded 
[gray] band representing the 90\% confidence level of the fit, with the 
resonances and total background components of the fit represented by 
dashed [blue] and dotted [red] lines, respectively.

The combinatorial background contribution is extracted using event 
mixing techniques and is normalized to match the like-sign background.  
To determine systematic uncertainties on the relative yields that vary 
from 1\% in $p$$+$$p$ collisions to 4\% in \heau collisions, the mass 
range over which the normalization is done is varied from a nominal 
range of 2--5 GeV/$c^2$ to 1.5--5 GeV/$c^2$ and 2.5--5.5 GeV/$c^2$. The 
correlated background is modeled in the fit by an exponential.  In the 
fitting procedure, the combinatorial background contribution is fixed 
using the methods previously described, while the shape and 
normalization parameters of the correlated background are allowed to 
vary.

\begin{table*}[tbh]
\caption{Summary of the measured ratios of $\psi(2S)/\psi(1S)$ mesons.  
The first (second) values in the rightmost column represent 
statistical (systematic) uncertainties.}
\begin{ruledtabular} \begin{tabular}{clccc}
  Collision system & Rapidity interval & 
$N^{\psi(1S)}_{\rm meas} $ & $N^{\psi(2S)}_{\rm meas} $ & $ 
\frac{B{^{\psi(2S)}_{\mu\mu}} \sigma_{\psi(2S)}}{B{^{\psi(1S)}_{\mu\mu}} 
\sigma_{\psi(1S)}}$ (\%) \\ \hline
  $p$$+$$p$      & $1.2<|y|<2.2$    & 17120$\pm$392  
& 519$\pm$51             & 2.43$\pm$0.18$\pm$0.29\\
  $p+$Al         & $1.2<y<2.2$ ($p$-going)   
& 1497$\pm$142           & 52$\pm$11 & 2.73$\pm$0.64$\pm$0.13\\
  $p+$Al         &  $-2.2<y<-1.2$ (Al-going)    
& 1463$\pm$109           &   25$\pm$11 & 1.37$\pm$0.61$\pm$0.16\\
  $p+$Au         &  $1.2<y<2.2$ ($p$-going)  
& 3893$\pm$147           &   117$\pm$18  & 2.38$\pm$0.37$\pm$0.30\\
  $p+$Au         & $-2.2<y<-1.2$ (Au-going) 
& 3561$\pm$180           &   51$\pm$18  & 1.16$\pm$0.42$\pm$0.17\\
  $^{3}$He$+$Au  & $1.2<y<2.2$ ($^{3}$He-going) 
& 959$\pm$64             &   27$\pm$9.3 & 2.24$\pm$0.78$\pm$0.32\\
  $^{3}$He$+$Au  & $-2.2<y<-1.2$ (Au-going) 
& 1772$\pm$132           &   35$\pm$15   & 1.59$\pm$0.67$\pm$0.60\\
\end{tabular} \end{ruledtabular}
\label{tab:summary_table}
\end{table*}

The resonances are represented in the fit by the sum of a crystal-ball 
function~\cite{Skwarnicki:1986xj} plus a Gaussian.  The crystal ball is 
a continuous piecewise function that is comprised of a Gaussian on the 
high mass side and an exponential on the low mass side, which accounts 
for tails due to muon energy loss straggling in the absorbers.  Due to 
10 cm of additional steel absorber in the North arm as compared to the 
South, and the variations in the meson $p_{T}$ spectra in the different 
collision systems, the low mass tail on both peaks is expected to be 
different between the two arms and various systems.  Therefore the 
parameters describing this low-side tail are allowed to float during 
fitting. The additional Gaussian is needed to account for tracks which 
have fewer than 14 hits out of a possible 16 in the muon tracker and 
therefore form pairs with poorer mass resolution.  By analyzing pairs 
formed with one and two of these poorly reconstructed tracks, the width 
of the $\psi(1S)$ is found to vary from $\sim$200 MeV/$c^{2}$ to 
$\sim$250 MeV/$c^{2}$.  Therefore, the width of this second Gaussian 
under the $\psi(1S)$ is set to 200 MeV/$c^{2}$ and varied by $\pm$25\% 
to determine a systematic uncertainty on the relative yields.  From 
previous measurements and simulations~\cite{FVTX_TDR}, the width of the 
$\psi(2S)$ is expected to be 1.15 times the width of the $\psi(1S)$ 
peak; therefore the second Gaussian under this resonance is set to a 
nominal value of 230 MeV/$c^{2}$ and also varied by $\pm$25\%.  This is 
the dominant source of systematic uncertainty on this measurement, and 
ranges from 8\% in $p$$+$$p$ collisions to 37\% in the Au-going 
direction in \heau collisions, due to the larger combinatorial 
background.

The difference between the centers of the $\psi(2S)$ and $\psi(1S)$ 
peaks is set to the Particle Data Group value~\cite{Agashe:2014kda} of 
0.589 MeV/$c^{2}$, and the width of the $\psi(2S)$ crystal-ball function 
is set to 1.15 times the width of the $\psi(1S)$ peak, again following 
expectations of the mass resolution in the muon spectrometer.  This 
constraint is varied from 1.1 to 1.2 times the $\psi(1S)$ width to 
determine a systematic uncertainty on the relative yields of $\sim3$\% 
for all systems (variations outside this range do not converge on stable 
fit parameter values).  Table~\ref{tab:summary_table} gives a summary of 
the counts measured in each collision system.

The physics quantity of interest here is the ratio of of the cross 
sections $\sigma$ of the $\psi(2S)$ to $\psi(1S)$ mesons, multiplied by 
their respective branching ratio to dimuons $B_{\mu\mu}$.  The counts 
$N_{\rm meas}$ of each meson are determined by the fits to the dimuon mass 
spectra, and are corrected for the PHENIX muon spectrometer pair 
acceptance $\Omega$, pair detection efficiency $\epsilon$, and dimuon 
trigger efficiency $\epsilon^{\rm trig}$ for each species.  The ratio of 
$\psi(2S)$ to $\psi(1S)$ mesons is given by

\begin{equation}
\frac{B{^{\psi(2S)}_{\mu\mu}} 
\sigma_{\psi(2S)}}{B{^{\psi(1S)}_{\mu\mu}} \sigma_{\psi(1S)}} = 
\frac{N^{\psi(2S)}_{\rm meas}}{N^{\psi(1S)}_{\rm meas}} 
\frac{\Omega_{\psi(1S)}\epsilon_{\psi(1S)}}
{\Omega_{\psi(2S)}\epsilon_{\psi(2S)}} 
\frac{\epsilon^{\rm trig}_{\psi(1S)}}{\epsilon^{\rm trig}_{\psi(2S)}}
\label{eqn:2s1sfrac}
\end{equation}

The acceptance$\times$efficiency factor 
$\frac{\Omega_{\psi(1S)}\epsilon_{\psi(1S)}}
{\Omega_{\psi(2S)}\epsilon_{\psi(2S)}}$ 
is determined via a full {\sc geant}4 simulation of the PHENIX 
detector~\cite{Agostinelli:2002hh}.  A set of simulated dimuons, with a 
continuum of realistic mass, $p_{T}$, and rapidity distributions, is 
passed through the simulated PHENIX detector. The ratio of the 
acceptance$\times$efficiency at the $\psi(1S)$ mass value to the 
$\psi(2S)$ mass value is found to be $\sim$0.8 for each arm. Since the 
$p_{T}$ distributions of $\psi(2S)$ and $\psi(1S)$ mesons have not been 
measured in all the collision systems presented here, a systematic 
uncertainty on the acceptance$\times$efficiency factor is determined by 
changing the assumed $p_{T}$ spectra.  Different dimuon samples are 
prepared assuming the $\psi(1S)$ spectrum follows the distribution 
previously measured in $p+p$ collisions and $d+$Au collisions at forward 
and backward rapidity \cite{Adare:2012qf}, and $m_{T}$-scaling 
\cite{Adare:2010fe} these distributions to approximate the $\psi(2S)$ 
spectrum.  Adjusting the correction factors between these various 
assumptions gives a 2\% systematic uncertainty on the relative yields.

The relative dimuon trigger efficiency $\frac{\epsilon^{\rm 
trig}_{\psi(1S)}}{\epsilon^{\rm trig}_{\psi(2S)}}$ is measured by 
finding the proportion of dimuon pairs in the Minimum Bias triggered 
data set which also fire the dimuon trigger.  This small correction 
factor is $\sim$0.97, and a relative systematic uncertainty of 1\% on 
the relative yields is assigned due to the statistical uncertainties on 
the dimuon sample in the Minimum Bias data set.  Because no significant 
$\psi(1S)$ polarization has been measured at PHENIX~\cite{Adare:2009js}, 
all corrections are calculated under the assumption that the mesons are 
unpolarized.

\section{Results and Discussion}

\begin{figure}[htbp]
  \centering
  \includegraphics[width=1.0\linewidth]{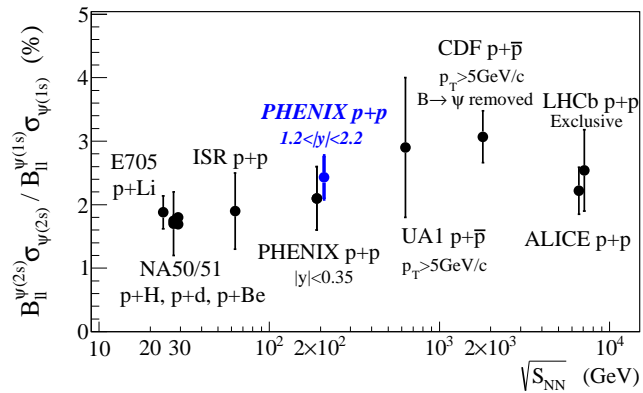}
\caption{
Comparison of world data on the ratio of $\psi(2S)/\psi(1S)$ mesons in 
dilepton decays~\cite{Antoniazzi:1992iv,Alessandro:2006jt,Abreu:1998rx,Clark:1978mg,Adare:2011vq,Albajar:1990hf,Abe:1997jz,Aaij:2013jxj,Abelev:2014qha}.
}
  \label{fig:psip_jpsi_sqrts}
\end{figure}

The $p_{T}$ integrated ratios of $\psi(2S)$ to $\psi(1S)$ mesons 
extracted from the North and South spectrometer arms in the symmetric 
$p$$+$$p$ collision system agree within two standard deviations.  These 
data points are averaged with a weighted least squared procedure, using 
the inverse of the square of the statistical uncertainties as weights 
(the same procedure that is used by the Particle Data Group to combine 
measurements of the same quantity \cite{Agashe:2014kda}).  The resulting 
data point is shown in Fig.~\ref{fig:psip_jpsi_sqrts}, and is consistent 
with world data taken at other collision energies.  Little difference is 
observed in the ratio of $\psi(2S)$ to $\psi(1S)$ mesons produced in 
collisions with center-of-mass energies that range over nearly three 
orders of magnitude, within uncertainties.  This may imply that despite 
significant differences in the total charm cross section across these 
energies, once a precursor $c\bar{c}$ is produced, the probability that 
it will project onto a given charmonium state is insensitive to the 
conditions which formed the initial $c\bar{c}$ pair.

\begin{figure}[htbp]
  \centering
  \includegraphics[width=1.0\linewidth]{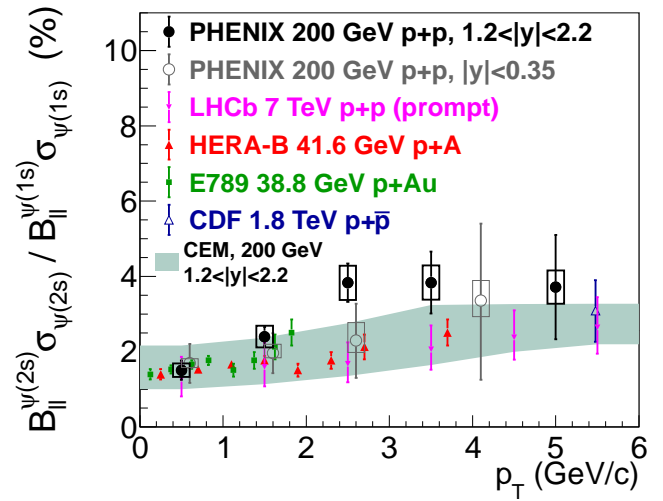}
  \caption{
Comparison of world data on the ratio of $\psi(2S)/\psi(1S)$ 
mesons as a function of 
$p_{T}$~\cite{Adare:2011vq, Aaij:2012ag,Abt:2006va, Schub:1995pu,Abe:1997jz}, 
along with a calculation from a color evaporation model (CEM) at 
$\sqs$=200 GeV~\cite{Ramona_privcomm}.
}
  \label{fig:pp_pt_CEM_world}
\end{figure}

The same ratio is shown as a function of $p_{T}$ in 
Fig.~\ref{fig:pp_pt_CEM_world}, along with a calculation based on a 
modified color evaporation model of charmonium production at 
200~GeV~\cite{Ramona_privcomm}.  This model factorizes the initial 
production of the $c\bar{c}$ pair from the color neutralization process 
via emission of soft gluons.  The ratios reported here are somewhat 
higher than the model for $p_{T}>$ 2 GeV/$c$, but the data's limited 
statistical precision and significant theoretical error bands preclude 
any firm conclusions.

\begin{figure}[htbp]
  \centering
  \includegraphics[width=1.0\linewidth]{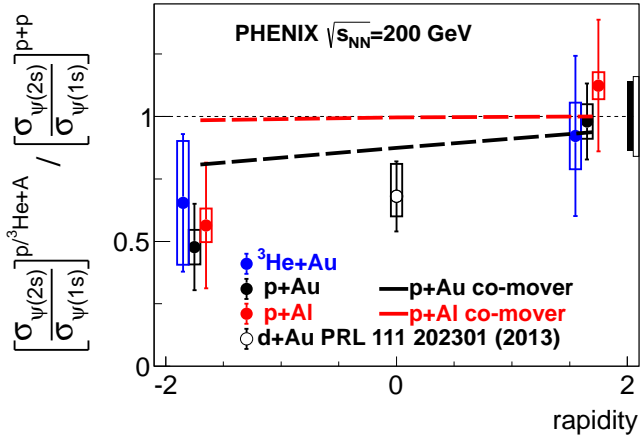}
\caption{The double ratio of $\psi(2S)/\psi(1S)$ mesons measured in in 
$p/d/^{3}$He$+A$ collisions to that same ratio in $p$+$p$ collisions at 
\sqs=200 GeV, with a calculation based on breakup by comoving 
particles~\cite{Ferreiro_privcomm}.  The bars (boxes) on the data points 
represent the statistical (systematic) uncertainties, and the shaded (open) 
box around unity represents the global uncertainty on the forward/backward 
(mid) rapidity data. 
}
  \label{fig:double_rat_corrected}
\end{figure}

From Fig.~\ref{fig:fits} it is apparent that the $\psi(2S)$ peaks are 
suppressed relative to the $\psi(1S)$ peak in the columns on the left 
(in the $A$-going direction).  Quantitative comparisons are accomplished 
by calculating the double ratio of $\psi(2S)/\psi(1S)$ production in 
$p/^{3}$He$+$$A$ collisions to the ratio found in $p$$+$$p$ collisions, 
as shown in Fig.~\ref{fig:double_rat_corrected}.  A previously published 
data point from midrapidity $d+$Au collisions at the same energy is also 
included for comparison~\cite{Adare:2013ezl}.  We see that at forward 
rapidity, the double ratio is consistent with unity in all three 
collision systems, indicating that any possible nuclear effects on the 
two charmonium states are comparable.  Because these states are not 
fully formed until after they exit the nucleus, the fact that any 
nuclear effects have an equal magnitude on both states suggests that 
there are no significant final state effects on the pair which occur in 
this rapidity region.

At backward rapidity, the ratios in all collision systems are suppressed 
by a factor of $\sim$2.  The mechanism for this preferential suppression 
of the $\psi(2S)$ relative to the $\psi(1S)$ is expected to occur after 
the $\psi(1S)$ formation time.  A significant difference in the late 
stages of the collision between this region and forward rapidity is the 
presence of a larger number of comoving hadrons (see~\cite{Alver:2010ck} 
for measurements of the charged particle rapidity distributions in 
$d+$Au, a similar collision system). Once the $c\bar{c}$ pair exits the 
nucleus, it may be subject to interactions with these particles that can 
lead to breakup of the charmonium state, which are expected to be more 
pronounced on the less tightly bound $\psi(2S)$.  While the exact nature 
of the interactions is not well understood, a model based on an 
absorption cross section that depends on the pair binding energy shows a 
preferential suppression that increases from forward to backward 
rapidity with comoving hadron multiplicity~\cite{Ferreiro_privcomm}.  
The model shows a similar trend with the $p+$Au data, although it 
underestimates the relative suppression at backward rapidity (see the 
solid [black] line in Fig.~\ref{fig:double_rat_corrected}).  However, in 
\pal, the model predicts almost no relative suppression at backward 
rapidity (dotted [red] line), while the data show a relative suppression 
similar to that in \pau, within significant uncertainties.

Measurements in 5.02 TeV $p+$Pb collisions at the Large Hadron Collider 
(LHC) have also observed that the $\psi(2S)$ is preferentially 
suppressed compared to the 
$\psi(1S)$~\cite{Abelev:2014zpa,Aaij:2016eyl}.  However, at this 
collision energy, the preferential suppression shows no significant 
rapidity dependence.  This may indicate that the nuclear effects which 
preferentially suppress the $\psi(2S)$ are similar at forward and 
backward rapidity at the LHC.  The charged particle pseudorapidity 
density per participant nucleon $(dN_{ch}/d\eta)/<N_{\rm part}>$ in 
$p$$+$Pb collisions at the LHC was found to be almost twice as high as 
$d+$Au collisions at RHIC~\cite{ALICE:2012xs}, so interactions with 
comoving particles may be more significant, even at forward rapidity.

 \begin{table*}[bth]
\caption{
The charged particle multiplicity $dN_{ch}/d\eta$ determined from AMPT 
simulations~\cite{Lin:2002gc,Lin:2004en} and the nuclear overlap $S_{T}$ 
from Glauber simulations~\cite{Loizides:2014vua,Nagle:2013lja} for 
the data shown in Fig.~\protect\ref{fig:double_rat_Nch}. Here we assume 
$dN/d\eta=\frac{3}{2}dN_{ch}/d\eta$, see text.
}
\begin{ruledtabular} \begin{tabular}{cccccc}
 Collision system & $\sqrt{s_{_{NN}}}$   &      Rapidity interval        
  & $(dN_{ch}/d\eta)_{AMPT} $ & $<S_{T}>$  (fm$^2$)  
  & $\frac{(dN/d\eta)_{AMPT}}{ <S_{T}>}$  (fm$^{-2}$) \\ \hline
 $p+$Al   & 200 GeV  & $-2.2<y< -1.2$   &  5.9 & 0.82 & 10.8 \\
 $p+$Al   & 200 GeV  & $1.2<y<2.2$   &  4.8 & 0.82  & 8.8\\
 $p+$Au   & 200 GeV  & $-2.2<y< -1.2$  & 10.2 & 1.01 &  15.1\\
 $p+$Au   & 200 GeV  & $1.2<y<2.2$  &  6.5 & 1.01 & 9.7\\
 $^{3}$He+Au  & 200 GeV  & $-2.2<y< -1.2$  & 20.3 & 2.35 & 13.0\\
 $^{3}$He+Au  & 200 GeV  & $1.2<y<2.2$   & 13.5 & 2.35&  8.6\\
 $p+$Pb   & 5.02 TeV & $-4.46<y<-2.03$ & 19.9 &  1.3 & 23.0\\
 $p+$Pb   & 5.02 TeV & $2.03<y<3.53$ &   15 &  1.3 & 17.3\\
 $p+$Pb   & 5.02 TeV & $-4.0<y<-2.5$   & 20.4 &  1.3 & 23.5\\
 $p+$Pb   & 5.02 TeV & $2.5<y<4.0$   & 13.7 &  1.3 & 15.8\\
\end{tabular} \end{ruledtabular}
\label{tab:N_table}
\end{table*}

\begin{figure}[htbp]
  \centering
  \includegraphics[width=1.0\linewidth]{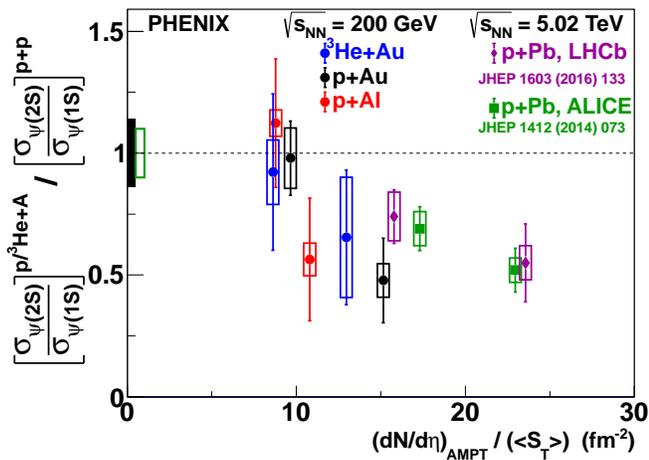}
  \caption{The double ratio of $\psi(2S)/\psi(1S)$ mesons measured in in 
$p/^{3}$He$+$$A$ collisions to that same ratio in $p$$+$$p$ collisions as a 
function of comoving particle density.  The shaded [black] (open [green]) 
box around unity represents the global uncertainty on the PHENIX (ALICE) 
data.}
  \label{fig:double_rat_Nch}
\end{figure}

For a direct comparison, the double ratios measured by PHENIX and at the 
LHC are plotted together in Fig.~\ref{fig:double_rat_Nch} as a function 
of comoving particle density, which is defined as the particle 
multiplicity $dN/d\eta$ evaluated over the same rapidity interval as the 
charmonia measurement, divided by the nuclear overlap $<S_{T}>.$ As 
measurements of the particle multiplicity do not exist for all these 
systems, the charged particle multiplicity $dN_{ch}/d\eta$ is determined 
by AMPT simulations~\cite{Lin:2002gc,Lin:2004en} and multiplied by 3/2 
to give the approximate total particle multiplicity $dN/d\eta$.  The 
nuclear overlap $<S_{T}>$ is defined as

\begin{equation}
\textit{$<S_{T}> = 4\pi \sqrt{<x^2><y^2> - <xy>^2}$}
\label{eqn:other2s1sfrac}
\end{equation}

\noindent where $x$ and $y$ are the spatial coordinates of the 
participating nucleons.  This quantity is found via Monte Carlo Glauber 
simulations of the various collision species~\cite{Loizides:2014vua}, 
with the $^{3}$He geometry modified as described in 
Ref.~\cite{Nagle:2013lja}.  Table~\ref{tab:N_table} summarizes the 
results of these simulations.

Figure~\ref{fig:double_rat_Nch} shows that the double ratio decreases as 
the comoving particle density increases, which is qualitatively 
consistent with expectations of charmonium breakup through final state 
interactions.  The forward-rapidity PHENIX data at relatively low 
comover density shows no preferential suppression of 
$\psi(2S)/\psi(1S)$, but the forward rapidity LHC data show a relative 
suppression that is comparable to the backward rapidity PHENIX data at 
similar comover density, within uncertainties.  The backward LHC data, 
at the highest comover density, also shows relative suppression.

Understanding suppression due to comovers could play a critical role in 
interpreting quarkonia data from $A$$+$$A$ collisions.  Existing data on 
excited charmonia states in $A$$+$$A$ collisions generally suffer from 
poor statistics, large combinatorial backgrounds, and cover limited 
$p_{T}$ ranges~\cite{Adam:2015isa,Khachatryan:2014bva}.  However, a 
clear sequential suppression of the excited bottomonium states has been 
observed in Pb$+$Pb collisions at the LHC~\cite{Chatrchyan:2012lxa}.  
While color screening is expected to play a role, given the high charged 
particle density in these collisions, it is reasonable to expect that 
similar breakup mechanisms can also have an effect on these states.  In 
particular, the highly suppressed $\Upsilon(3s)$ state has a binding 
energy of only $\sim$200 MeV, and may be especially sensitive to breakup 
through interactions with comoving particles.
 
\section{Conclusions}
 
In conclusion, we have found that the relative production of $\psi(2S)$ 
to $\psi(1S)$ mesons in \pp collisions at \sqs = 200 GeV is consistent 
with expectations from a modified color evaporation model of charmonium 
production.  In $p/^{3}$He$+A$ collisions at forward rapidity we observe 
no difference in the $\psi(2S)/\psi(1S)$ ratio relative to $p$$+$$p$ 
collisions, which indicates that any possible nuclear effects that are 
present in this rapidity region are common between the two states, and 
therefore appear to occur on a timescale that is short compared to the 
charmonium formation time.  At backward rapidity, where the comoving 
particle density is higher, we find that the $\psi(2S)$ is 
preferentially suppressed by a factor of $\sim2$.  This effect is likely 
not due to any interaction in the nucleus, because the $c\bar{c}$ pair 
exits the nucleus before final meson formation occurs. The preferential 
suppression appears consistent with interactions of the fully formed 
color-neutral meson with comoving particles.  The magnitude of this 
breakup mechanism is dependent on the meson binding energy, and is 
likely important for interpretation of sequential screening of quarkonia 
in $A$$+$$A$ collisions.

\section*{Acknowledgments}   

We thank the staff of the Collider-Accelerator and Physics
Departments at Brookhaven National Laboratory and the staff of
the other PHENIX participating institutions for their vital
contributions.  
We also thank Elena Ferreiro and Ramona Vogt for providing model 
calculations.  
We acknowledge support from the Office of Nuclear Physics in the
Office of Science of the Department of Energy,
the National Science Foundation, 
Abilene Christian University Research Council, 
Research Foundation of SUNY, and
Dean of the College of Arts and Sciences, Vanderbilt University 
(U.S.A),
Ministry of Education, Culture, Sports, Science, and Technology
and the Japan Society for the Promotion of Science (Japan),
Conselho Nacional de Desenvolvimento Cient\'{\i}fico e
Tecnol{\'o}gico and Funda\c c{\~a}o de Amparo {\`a} Pesquisa do
Estado de S{\~a}o Paulo (Brazil),
Natural Science Foundation of China (P.~R.~China),
Croatian Science Foundation and
Ministry of Science, Education, and Sports (Croatia),
Ministry of Education, Youth and Sports (Czech Republic),
Centre National de la Recherche Scientifique, Commissariat
{\`a} l'{\'E}nergie Atomique, and Institut National de Physique
Nucl{\'e}aire et de Physique des Particules (France),
Bundesministerium f\"ur Bildung und Forschung, Deutscher
Akademischer Austausch Dienst, and Alexander von Humboldt Stiftung (Germany),
National Science Fund, OTKA, K\'aroly R\'obert University College, 
and the Ch. Simonyi Fund (Hungary),
Department of Atomic Energy and Department of Science and Technology (India), 
Israel Science Foundation (Israel), 
Basic Science Research Program through NRF of the Ministry of Education (Korea),
Physics Department, Lahore University of Management Sciences (Pakistan),
Ministry of Education and Science, Russian Academy of Sciences,
Federal Agency of Atomic Energy (Russia),
VR and Wallenberg Foundation (Sweden), 
the U.S. Civilian Research and Development Foundation for the
Independent States of the Former Soviet Union, 
the Hungarian American Enterprise Scholarship Fund,
and the US-Israel Binational Science Foundation.



\begin{thebibliography}{57}%
\makeatletter
\providecommand \@ifxundefined [1]{%
 \@ifx{#1\undefined}
}%
\providecommand \@ifnum [1]{%
 \ifnum #1\expandafter \@firstoftwo
 \else \expandafter \@secondoftwo
 \fi
}%
\providecommand \@ifx [1]{%
 \ifx #1\expandafter \@firstoftwo
 \else \expandafter \@secondoftwo
 \fi
}%
\providecommand \natexlab [1]{#1}%
\providecommand \enquote  [1]{``#1''}%
\providecommand \bibnamefont  [1]{#1}%
\providecommand \bibfnamefont [1]{#1}%
\providecommand \citenamefont [1]{#1}%
\providecommand \href@noop [0]{\@secondoftwo}%
\providecommand \href [0]{\begingroup \@sanitize@url \@href}%
\providecommand \@href[1]{\@@startlink{#1}\@@href}%
\providecommand \@@href[1]{\endgroup#1\@@endlink}%
\providecommand \@sanitize@url [0]{\catcode `\\12\catcode `\$12\catcode
  `\&12\catcode `\#12\catcode `\^12\catcode `\_12\catcode `\%12\relax}%
\providecommand \@@startlink[1]{}%
\providecommand \@@endlink[0]{}%
\providecommand \url  [0]{\begingroup\@sanitize@url \@url }%
\providecommand \@url [1]{\endgroup\@href {#1}{\urlprefix }}%
\providecommand \urlprefix  [0]{URL }%
\providecommand \Eprint [0]{\href }%
\providecommand \doibase [0]{http://dx.doi.org/}%
\providecommand \selectlanguage [0]{\@gobble}%
\providecommand \bibinfo  [0]{\@secondoftwo}%
\providecommand \bibfield  [0]{\@secondoftwo}%
\providecommand \translation [1]{[#1]}%
\providecommand \BibitemOpen [0]{}%
\providecommand \bibitemStop [0]{}%
\providecommand \bibitemNoStop [0]{.\EOS\space}%
\providecommand \EOS [0]{\spacefactor3000\relax}%
\providecommand \BibitemShut  [1]{\csname bibitem#1\endcsname}%
\let\auto@bib@innerbib\@empty
\bibitem [{\citenamefont {Matsui}\ and\ \citenamefont
  {Satz}(1986)}]{Matsui:1986dk}%
  \BibitemOpen
  \bibfield  {author} {\bibinfo {author} {\bibfnamefont {T.}~\bibnamefont
  {Matsui}}\ and\ \bibinfo {author} {\bibfnamefont {H.}~\bibnamefont {Satz}},\
  }\bibfield  {title} {\enquote {\bibinfo {title} {{$J/\psi$ Suppression by
  Quark-Gluon Plasma Formation}},}\ }\href@noop {} {\bibfield  {journal}
  {\bibinfo  {journal} {Phys. Lett. B}\ }\textbf {\bibinfo {volume} {178}},\
  \bibinfo {pages} {416} (\bibinfo {year} {1986})}\BibitemShut {NoStop}%
\bibitem [{\citenamefont {Cacciari}\ \emph {et~al.}(2005)\citenamefont
  {Cacciari}, \citenamefont {Nason},\ and\ \citenamefont
  {Vogt}}]{Cacciari:2005rk}%
  \BibitemOpen
  \bibfield  {author} {\bibinfo {author} {\bibfnamefont {M.}~\bibnamefont
  {Cacciari}}, \bibinfo {author} {\bibfnamefont {P.}~\bibnamefont {Nason}}, \
  and\ \bibinfo {author} {\bibfnamefont {R.}~\bibnamefont {Vogt}},\ }\bibfield
  {title} {\enquote {\bibinfo {title} {{QCD predictions for charm and bottom
  production at RHIC}},}\ }\href@noop {} {\bibfield  {journal} {\bibinfo
  {journal} {Phys. Rev. Lett.}\ }\textbf {\bibinfo {volume} {95}},\ \bibinfo
  {pages} {122001} (\bibinfo {year} {2005})}\BibitemShut {NoStop}%
\bibitem [{\citenamefont {Fritzsch}(1977)}]{Fritzsch:1977ay}%
  \BibitemOpen
  \bibfield  {author} {\bibinfo {author} {\bibfnamefont {H.}~\bibnamefont
  {Fritzsch}},\ }\bibfield  {title} {\enquote {\bibinfo {title} {{Producing
  Heavy Quark Flavors in Hadronic Collisions: A Test of Quantum
  Chromodynamics}},}\ }\href@noop {} {\bibfield  {journal} {\bibinfo  {journal}
  {Phys. Lett. B}\ }\textbf {\bibinfo {volume} {67}},\ \bibinfo {pages}
  {217--221} (\bibinfo {year} {1977})}\BibitemShut {NoStop}%
\bibitem [{\citenamefont {Bodwin}\ \emph {et~al.}(2005)\citenamefont {Bodwin},
  \citenamefont {Braaten},\ and\ \citenamefont {Lee}}]{Bodwin:2005hm}%
  \BibitemOpen
  \bibfield  {author} {\bibinfo {author} {\bibfnamefont {G.~T.}\ \bibnamefont
  {Bodwin}}, \bibinfo {author} {\bibfnamefont {E.}~\bibnamefont {Braaten}}, \
  and\ \bibinfo {author} {\bibfnamefont {J.}~\bibnamefont {Lee}},\ }\bibfield
  {title} {\enquote {\bibinfo {title} {{Comparison of the color-evaporation
  model and the NRQCD factorization approach in charmonium production}},}\
  }\href {\doibase 10.1103/PhysRevD.72.014004} {\bibfield  {journal} {\bibinfo
  {journal} {Phys. Rev. D}\ }\textbf {\bibinfo {volume} {72}},\ \bibinfo
  {pages} {014004} (\bibinfo {year} {2005})}\BibitemShut {NoStop}%
\bibitem [{\citenamefont {Kang}\ \emph {et~al.}(2014)\citenamefont {Kang},
  \citenamefont {Ma}, \citenamefont {Qiu},\ and\ \citenamefont
  {Sterman}}]{Kang:2014tta}%
  \BibitemOpen
  \bibfield  {author} {\bibinfo {author} {\bibfnamefont {Z.~B.}\ \bibnamefont
  {Kang}}, \bibinfo {author} {\bibfnamefont {Y.~Q.}\ \bibnamefont {Ma}},
  \bibinfo {author} {\bibfnamefont {J.~W.}\ \bibnamefont {Qiu}}, \ and\
  \bibinfo {author} {\bibfnamefont {G.}~\bibnamefont {Sterman}},\ }\bibfield
  {title} {\enquote {\bibinfo {title} {{Heavy Quarkonium Production at Collider
  Energies: Factorization and Evolution}},}\ }\href {\doibase
  10.1103/PhysRevD.90.034006} {\bibfield  {journal} {\bibinfo  {journal} {Phys.
  Rev. D}\ }\textbf {\bibinfo {volume} {90}},\ \bibinfo {pages} {034006}
  (\bibinfo {year} {2014})}\BibitemShut {NoStop}%
\bibitem [{\citenamefont {Helenius}\ \emph {et~al.}()\citenamefont {Helenius},
  \citenamefont {Eskola}, \citenamefont {Honkanen},\ and\ \citenamefont
  {Salgado}}]{Helenius:2012wd}%
  \BibitemOpen
  \bibfield  {author} {\bibinfo {author} {\bibfnamefont {I.}~\bibnamefont
  {Helenius}}, \bibinfo {author} {\bibfnamefont {K.~J.}\ \bibnamefont
  {Eskola}}, \bibinfo {author} {\bibfnamefont {H.}~\bibnamefont {Honkanen}}, \
  and\ \bibinfo {author} {\bibfnamefont {C.~A.}\ \bibnamefont {Salgado}},\
  }\href@noop {} {\enquote {\bibinfo {title} {{Impact-Parameter Dependent
  Nuclear Parton Distribution Functions: EPS09s and EKS98s and Their
  Applications in Nuclear Hard Processes}},}\ }\bibinfo {note} {{J. High Energy
  Phys. {\bf 07 (2012)} 073}}\BibitemShut {NoStop}%
\bibitem [{\citenamefont {Vitev}(2007)}]{Vitev:2007ve}%
  \BibitemOpen
  \bibfield  {author} {\bibinfo {author} {\bibfnamefont {I.}~\bibnamefont
  {Vitev}},\ }\bibfield  {title} {\enquote {\bibinfo {title} {{Non-Abelian
  energy loss in cold nuclear matter}},}\ }\href {\doibase
  10.1103/PhysRevC.75.064906} {\bibfield  {journal} {\bibinfo  {journal} {Phys.
  Rev. C}\ }\textbf {\bibinfo {volume} {75}},\ \bibinfo {pages} {064906}
  (\bibinfo {year} {2007})}\BibitemShut {NoStop}%
\bibitem [{\citenamefont {Vogt}(2003)}]{Vogt:2001nh}%
  \BibitemOpen
  \bibfield  {author} {\bibinfo {author} {\bibfnamefont {R.}~\bibnamefont
  {Vogt}},\ }\bibfield  {title} {\enquote {\bibinfo {title} {{The $A$
  dependence of open charm and bottom production}},}\ }\href {\doibase
  10.1142/S0218301303001272} {\bibfield  {journal} {\bibinfo  {journal} {Int.
  J. Mod. Phys. E}\ }\textbf {\bibinfo {volume} {12}},\ \bibinfo {pages}
  {211--270} (\bibinfo {year} {2003})}\BibitemShut {NoStop}%
\bibitem [{\citenamefont {Adare}\ \emph
  {et~al.}(2013{\natexlab{a}})\citenamefont {Adare} \emph
  {et~al.}}]{Adare:2013piz}%
  \BibitemOpen
  \bibfield  {author} {\bibinfo {author} {\bibfnamefont {A.}~\bibnamefont
  {Adare}} \emph {et~al.} (\bibinfo {collaboration} {PHENIX Collaboration}),\
  }\bibfield  {title} {\enquote {\bibinfo {title} {{Quadrupole Anisotropy in
  Dihadron Azimuthal Correlations in Central $d$$+$Au Collisions at
  $\sqrt{s_{_{NN}}}=200$ GeV}},}\ }\href {\doibase
  10.1103/PhysRevLett.111.212301} {\bibfield  {journal} {\bibinfo  {journal}
  {Phys. Rev. Lett.}\ }\textbf {\bibinfo {volume} {111}},\ \bibinfo {pages}
  {212301} (\bibinfo {year} {2013}{\natexlab{a}})}\BibitemShut {NoStop}%
\bibitem [{\citenamefont {Adare}\ \emph
  {et~al.}(2015{\natexlab{a}})\citenamefont {Adare} \emph
  {et~al.}}]{Adare:2014keg}%
  \BibitemOpen
  \bibfield  {author} {\bibinfo {author} {\bibfnamefont {A.}~\bibnamefont
  {Adare}} \emph {et~al.} (\bibinfo {collaboration} {PHENIX Collaboration}),\
  }\bibfield  {title} {\enquote {\bibinfo {title} {{Measurement of long-range
  angular correlation and quadrupole anisotropy of pions and (anti)protons in
  central $d$$+$Au collisions at $\sqrt{s_{_{NN}}}=200$ GeV}},}\ }\href
  {\doibase 10.1103/PhysRevLett.114.192301} {\bibfield  {journal} {\bibinfo
  {journal} {Phys. Rev. Lett.}\ }\textbf {\bibinfo {volume} {114}},\ \bibinfo
  {pages} {192301} (\bibinfo {year} {2015}{\natexlab{a}})}\BibitemShut
  {NoStop}%
\bibitem [{\citenamefont {Adare}\ \emph
  {et~al.}(2015{\natexlab{b}})\citenamefont {Adare} \emph
  {et~al.}}]{Adare:2015ctn}%
  \BibitemOpen
  \bibfield  {author} {\bibinfo {author} {\bibfnamefont {A.}~\bibnamefont
  {Adare}} \emph {et~al.} (\bibinfo {collaboration} {PHENIX Collaboration}),\
  }\bibfield  {title} {\enquote {\bibinfo {title} {{Measurements of elliptic
  and triangular flow in high-multiplicity $^{3}$He$+$Au collisions at
  $\sqrt{s_{_{NN}}}=200$ GeV}},}\ }\href {\doibase
  10.1103/PhysRevLett.115.142301} {\bibfield  {journal} {\bibinfo  {journal}
  {Phys. Rev. Lett.}\ }\textbf {\bibinfo {volume} {115}},\ \bibinfo {pages}
  {142301} (\bibinfo {year} {2015}{\natexlab{b}})}\BibitemShut {NoStop}%
\bibitem [{\citenamefont {Du}\ and\ \citenamefont {Rapp}(2015)}]{Du:2015wha}%
  \BibitemOpen
  \bibfield  {author} {\bibinfo {author} {\bibfnamefont {X.}~\bibnamefont
  {Du}}\ and\ \bibinfo {author} {\bibfnamefont {R.}~\bibnamefont {Rapp}},\
  }\bibfield  {title} {\enquote {\bibinfo {title} {{Sequential Regeneration of
  Charmonia in Heavy-Ion Collisions}},}\ }\href {\doibase
  10.1016/j.nuclphysa.2015.09.006} {\bibfield  {journal} {\bibinfo  {journal}
  {Nucl. Phys. A}\ }\textbf {\bibinfo {volume} {943}},\ \bibinfo {pages}
  {147--158} (\bibinfo {year} {2015})}\BibitemShut {NoStop}%
\bibitem [{\citenamefont {Beraudo}\ \emph {et~al.}()\citenamefont {Beraudo}
  \emph {et~al.}}]{Beraudo:2015wsd}%
  \BibitemOpen
  \bibfield  {author} {\bibinfo {author} {\bibfnamefont {A.}~\bibnamefont
  {Beraudo}} \emph {et~al.},\ }\href@noop {} {\enquote {\bibinfo {title}
  {{Heavy-flavour production in high-energy $d$-Au and $p$-Pb collisions}},}\
  }\bibinfo {note} {{J. High Energy Phys. {\bf 03 (2016)} 123}}\BibitemShut
  {NoStop}%
\bibitem [{\citenamefont {Mocsy}\ \emph {et~al.}(2013)\citenamefont {Mocsy},
  \citenamefont {Petreczky},\ and\ \citenamefont {Strickland}}]{Mocsy:2013syh}%
  \BibitemOpen
  \bibfield  {author} {\bibinfo {author} {\bibfnamefont {A.}~\bibnamefont
  {Mocsy}}, \bibinfo {author} {\bibfnamefont {P.}~\bibnamefont {Petreczky}}, \
  and\ \bibinfo {author} {\bibfnamefont {M.}~\bibnamefont {Strickland}},\
  }\bibfield  {title} {\enquote {\bibinfo {title} {{Quarkonia in the Quark
  Gluon Plasma}},}\ }\href {\doibase 10.1142/S0217751X13400125} {\bibfield
  {journal} {\bibinfo  {journal} {Int. J. Mod. Phys. A}\ }\textbf {\bibinfo
  {volume} {28}},\ \bibinfo {pages} {1340012} (\bibinfo {year}
  {2013})}\BibitemShut {NoStop}%
\bibitem [{\citenamefont {Capella}\ \emph {et~al.}(1997)\citenamefont
  {Capella}, \citenamefont {Kaidalov}, \citenamefont {Kouider~Akil},\ and\
  \citenamefont {Gerschel}}]{Capella:1996va}%
  \BibitemOpen
  \bibfield  {author} {\bibinfo {author} {\bibfnamefont {A.}~\bibnamefont
  {Capella}}, \bibinfo {author} {\bibfnamefont {A.}~\bibnamefont {Kaidalov}},
  \bibinfo {author} {\bibfnamefont {A.}~\bibnamefont {Kouider~Akil}}, \ and\
  \bibinfo {author} {\bibfnamefont {C.}~\bibnamefont {Gerschel}},\ }\bibfield
  {title} {\enquote {\bibinfo {title} {{$J/\psi$ and $\psi^\prime$ suppression
  in heavy ion collisions}},}\ }\href {\doibase 10.1016/S0370-2693(96)01650-4}
  {\bibfield  {journal} {\bibinfo  {journal} {Phys. Lett. B}\ }\textbf
  {\bibinfo {volume} {393}},\ \bibinfo {pages} {431--436} (\bibinfo {year}
  {1997})}\BibitemShut {NoStop}%
\bibitem [{\citenamefont {Ferreiro}(2015)}]{Ferreiro:2014bia}%
  \BibitemOpen
  \bibfield  {author} {\bibinfo {author} {\bibfnamefont {E.~G.}\ \bibnamefont
  {Ferreiro}},\ }\bibfield  {title} {\enquote {\bibinfo {title} {{Excited
  charmonium suppression in proton-nucleus collisions as a consequence of
  comovers}},}\ }\href {\doibase 10.1016/j.physletb.2015.07.066} {\bibfield
  {journal} {\bibinfo  {journal} {Phys. Lett. B}\ }\textbf {\bibinfo {volume}
  {749}},\ \bibinfo {pages} {98--103} (\bibinfo {year} {2015})}\BibitemShut
  {NoStop}%
\bibitem [{\citenamefont {Satz}(2006)}]{Satz:2005hx}%
  \BibitemOpen
  \bibfield  {author} {\bibinfo {author} {\bibfnamefont {H.}~\bibnamefont
  {Satz}},\ }\bibfield  {title} {\enquote {\bibinfo {title} {{Colour
  deconfinement and quarkonium binding}},}\ }\href {\doibase
  10.1088/0954-3899/32/3/R01} {\bibfield  {journal} {\bibinfo  {journal} {J.
  Phys. G}\ }\textbf {\bibinfo {volume} {32}},\ \bibinfo {pages} {R25}
  (\bibinfo {year} {2006})}\BibitemShut {NoStop}%
\bibitem [{\citenamefont {Adare}\ \emph
  {et~al.}(2012{\natexlab{a}})\citenamefont {Adare} \emph
  {et~al.}}]{Adare:2012yxa}%
  \BibitemOpen
  \bibfield  {author} {\bibinfo {author} {\bibfnamefont {A.}~\bibnamefont
  {Adare}} \emph {et~al.} (\bibinfo {collaboration} {PHENIX Collaboration}),\
  }\bibfield  {title} {\enquote {\bibinfo {title} {{Cold-nuclear-matter effects
  on heavy-quark production in $d$+Au collisions at $\sqrt{s_{NN}}=200$
  GeV}},}\ }\href {\doibase 10.1103/PhysRevLett.109.242301} {\bibfield
  {journal} {\bibinfo  {journal} {Phys. Rev. Lett.}\ }\textbf {\bibinfo
  {volume} {109}},\ \bibinfo {pages} {242301} (\bibinfo {year}
  {2012}{\natexlab{a}})}\BibitemShut {NoStop}%
\bibitem [{\citenamefont {Adare}\ \emph {et~al.}(2014)\citenamefont {Adare}
  \emph {et~al.}}]{Adare:2013lkk}%
  \BibitemOpen
  \bibfield  {author} {\bibinfo {author} {\bibfnamefont {A.}~\bibnamefont
  {Adare}} \emph {et~al.} (\bibinfo {collaboration} {PHENIX Collaboration}),\
  }\bibfield  {title} {\enquote {\bibinfo {title} {{Cold-Nuclear-Matter Effects
  on Heavy-Quark Production at Forward and Backward Rapidity in $d$+Au
  Collisions at $\sqrt{s_{NN}}=200$GeV}},}\ }\href {\doibase
  10.1103/PhysRevLett.112.252301} {\bibfield  {journal} {\bibinfo  {journal}
  {Phys. Rev. Lett.}\ }\textbf {\bibinfo {volume} {112}},\ \bibinfo {pages}
  {252301} (\bibinfo {year} {2014})}\BibitemShut {NoStop}%
\bibitem [{\citenamefont {Leitch}\ \emph {et~al.}(2000)\citenamefont {Leitch}
  \emph {et~al.}}]{Leitch:1999ea}%
  \BibitemOpen
  \bibfield  {author} {\bibinfo {author} {\bibfnamefont {M.~J.}\ \bibnamefont
  {Leitch}} \emph {et~al.} (\bibinfo {collaboration} {NuSea Collaboration}),\
  }\bibfield  {title} {\enquote {\bibinfo {title} {{Measurement of $J/\psi$ and
  $\psi^\prime$ suppression in $p$-A collisions at 800-GeV/$c$}},}\ }\href
  {\doibase 10.1103/PhysRevLett.84.3256} {\bibfield  {journal} {\bibinfo
  {journal} {Phys. Rev. Lett.}\ }\textbf {\bibinfo {volume} {84}},\ \bibinfo
  {pages} {3256--3260} (\bibinfo {year} {2000})}\BibitemShut {NoStop}%
\bibitem [{\citenamefont {Alessandro}\ \emph {et~al.}(2006)\citenamefont
  {Alessandro} \emph {et~al.}}]{Alessandro:2006jt}%
  \BibitemOpen
  \bibfield  {author} {\bibinfo {author} {\bibfnamefont {B.}~\bibnamefont
  {Alessandro}} \emph {et~al.} (\bibinfo {collaboration} {NA50
  Collaboration}),\ }\bibfield  {title} {\enquote {\bibinfo {title} {{$J/\psi$
  and $\psi^\prime$ production and their normal nuclear absorption in
  proton-nucleus collisions at 400 GeV}},}\ }\href {\doibase
  10.1140/epjc/s10052-006-0079-4} {\bibfield  {journal} {\bibinfo  {journal}
  {Eur. Phys. J. C}\ }\textbf {\bibinfo {volume} {48}},\ \bibinfo {pages} {329}
  (\bibinfo {year} {2006})}\BibitemShut {NoStop}%
\bibitem [{\citenamefont {Arleo}\ \emph {et~al.}(2000)\citenamefont {Arleo},
  \citenamefont {Gossiaux}, \citenamefont {Gousset},\ and\ \citenamefont
  {Aichelin}}]{Arleo:1999af}%
  \BibitemOpen
  \bibfield  {author} {\bibinfo {author} {\bibfnamefont {F.}~\bibnamefont
  {Arleo}}, \bibinfo {author} {\bibfnamefont {P.~B.}\ \bibnamefont {Gossiaux}},
  \bibinfo {author} {\bibfnamefont {T.}~\bibnamefont {Gousset}}, \ and\
  \bibinfo {author} {\bibfnamefont {J.}~\bibnamefont {Aichelin}},\ }\bibfield
  {title} {\enquote {\bibinfo {title} {{Charmonium suppression in $p$-A
  collisions}},}\ }\href {\doibase 10.1103/PhysRevC.61.054906} {\bibfield
  {journal} {\bibinfo  {journal} {Phys. Rev. C}\ }\textbf {\bibinfo {volume}
  {61}},\ \bibinfo {pages} {054906} (\bibinfo {year} {2000})}\BibitemShut
  {NoStop}%
\bibitem [{\citenamefont {Adare}\ \emph
  {et~al.}(2013{\natexlab{b}})\citenamefont {Adare} \emph
  {et~al.}}]{Adare:2013ezl}%
  \BibitemOpen
  \bibfield  {author} {\bibinfo {author} {\bibfnamefont {A.}~\bibnamefont
  {Adare}} \emph {et~al.} (\bibinfo {collaboration} {PHENIX Collaboration}),\
  }\bibfield  {title} {\enquote {\bibinfo {title} {{Nuclear Modification of
  $\psi^\prime$, $\chi_{c}$ and $J/\psi$ Production in $d$+Au Collisions at
  $\sqrt{s_{NN}}=200$ GeV}},}\ }\href {\doibase 10.1103/PhysRevLett.111.202301}
  {\bibfield  {journal} {\bibinfo  {journal} {Phys. Rev. Lett.}\ }\textbf
  {\bibinfo {volume} {111}},\ \bibinfo {pages} {202301} (\bibinfo {year}
  {2013}{\natexlab{b}})}\BibitemShut {NoStop}%
\bibitem [{\citenamefont {McGlinchey}\ \emph {et~al.}(2013)\citenamefont
  {McGlinchey}, \citenamefont {Frawley},\ and\ \citenamefont
  {Vogt}}]{McGlinchey:2012bp}%
  \BibitemOpen
  \bibfield  {author} {\bibinfo {author} {\bibfnamefont {D.~C.}\ \bibnamefont
  {McGlinchey}}, \bibinfo {author} {\bibfnamefont {A.~D.}\ \bibnamefont
  {Frawley}}, \ and\ \bibinfo {author} {\bibfnamefont {R.}~\bibnamefont
  {Vogt}},\ }\bibfield  {title} {\enquote {\bibinfo {title} {{Impact parameter
  dependence of the nuclear modification of $J/\psi$ production in $d+$Au
  collisions at $\sqrt{s_{NN}}=200$ GeV}},}\ }\href {\doibase
  10.1103/PhysRevC.87.054910} {\bibfield  {journal} {\bibinfo  {journal} {Phys.
  Rev. C}\ }\textbf {\bibinfo {volume} {87}},\ \bibinfo {pages} {054910}
  (\bibinfo {year} {2013})}\BibitemShut {NoStop}%
\bibitem [{\citenamefont {Ma}\ and\ \citenamefont
  {Vogt}(2016)}]{Ramona_privcomm}%
  \BibitemOpen
  \bibfield  {author} {\bibinfo {author} {\bibfnamefont {Y.~Q.}\ \bibnamefont
  {Ma}}\ and\ \bibinfo {author} {\bibfnamefont {R.}~\bibnamefont {Vogt}},\
  }\href@noop {} {} (\bibinfo {year} {2016}),\ \bibinfo {note} {private
  communication and in preparation}\BibitemShut {NoStop}%
\bibitem [{\citenamefont {Akikawa}\ \emph {et~al.}(2003)\citenamefont {Akikawa}
  \emph {et~al.}}]{Akikawa:2003zs}%
  \BibitemOpen
  \bibfield  {author} {\bibinfo {author} {\bibfnamefont {H.}~\bibnamefont
  {Akikawa}} \emph {et~al.} (\bibinfo {collaboration} {PHENIX Collaboration}),\
  }\bibfield  {title} {\enquote {\bibinfo {title} {{PHENIX muon arms}},}\
  }\href {\doibase 10.1016/S0168-9002(02)01955-1} {\bibfield  {journal}
  {\bibinfo  {journal} {Nucl. Instrum. Methods Phys. Res., Sec. A}\ }\textbf
  {\bibinfo {volume} {499}},\ \bibinfo {pages} {537--548} (\bibinfo {year}
  {2003})}\BibitemShut {NoStop}%
\bibitem [{\citenamefont {Aidala}\ \emph {et~al.}(2014)\citenamefont {Aidala}
  \emph {et~al.}}]{Aidala:2013vna}%
  \BibitemOpen
  \bibfield  {author} {\bibinfo {author} {\bibfnamefont {C.}~\bibnamefont
  {Aidala}} \emph {et~al.},\ }\bibfield  {title} {\enquote {\bibinfo {title}
  {{The PHENIX Forward Silicon Vertex Detector}},}\ }\href {\doibase
  10.1016/j.nima.2014.04.017} {\bibfield  {journal} {\bibinfo  {journal} {Nucl.
  Instrum. Methods Phys. Res., Sec. A}\ }\textbf {\bibinfo {volume} {755}},\
  \bibinfo {pages} {44--61} (\bibinfo {year} {2014})}\BibitemShut {NoStop}%
\bibitem [{\citenamefont {Adare}\ \emph
  {et~al.}(2013{\natexlab{c}})\citenamefont {Adare} \emph
  {et~al.}}]{Adare:2012qf}%
  \BibitemOpen
  \bibfield  {author} {\bibinfo {author} {\bibfnamefont {A.}~\bibnamefont
  {Adare}} \emph {et~al.} (\bibinfo {collaboration} {PHENIX Collaboration}),\
  }\bibfield  {title} {\enquote {\bibinfo {title} {{Transverse-Momentum
  Dependence of the $J/\psi$ Nuclear Modification in $d+$Au Collisions at
  $\sqrt{s_{NN}}=200$ GeV}},}\ }\href {\doibase 10.1103/PhysRevC.87.034904}
  {\bibfield  {journal} {\bibinfo  {journal} {Phys. Rev. C}\ }\textbf {\bibinfo
  {volume} {87}},\ \bibinfo {pages} {034904} (\bibinfo {year}
  {2013}{\natexlab{c}})}\BibitemShut {NoStop}%
\bibitem [{\citenamefont {Adare}\ \emph
  {et~al.}(2012{\natexlab{b}})\citenamefont {Adare} \emph
  {et~al.}}]{Adare:2011vq}%
  \BibitemOpen
  \bibfield  {author} {\bibinfo {author} {\bibfnamefont {A.}~\bibnamefont
  {Adare}} \emph {et~al.} (\bibinfo {collaboration} {PHENIX Collaboration}),\
  }\bibfield  {title} {\enquote {\bibinfo {title} {{Ground and excited
  charmonium state production in $p+p$ collisions at $\sqrt{s}=200$ GeV}},}\
  }\href {\doibase 10.1103/PhysRevD.85.092004} {\bibfield  {journal} {\bibinfo
  {journal} {Phys. Rev. D}\ }\textbf {\bibinfo {volume} {85}},\ \bibinfo
  {pages} {092004} (\bibinfo {year} {2012}{\natexlab{b}})}\BibitemShut
  {NoStop}%
\bibitem [{\citenamefont {Skwarnicki}(1986)}]{Skwarnicki:1986xj}%
  \BibitemOpen
  \bibfield  {author} {\bibinfo {author} {\bibfnamefont {Tomasz}\ \bibnamefont
  {Skwarnicki}},\ }\emph {\bibinfo {title} {{A study of the radiative CASCADE
  transitions between the Upsilon-Prime and Upsilon resonances}}},\ \href@noop
  {} {Ph.D. thesis},\ \bibinfo  {school} {Cracow, INP} (\bibinfo {year}
  {1986})\BibitemShut {NoStop}%
\bibitem [{\citenamefont {Choudhury}\ \emph {et~al.}()\citenamefont {Choudhury}
  \emph {et~al.}}]{FVTX_TDR}%
  \BibitemOpen
  \bibfield  {author} {\bibinfo {author} {\bibfnamefont {R.~K.}\ \bibnamefont
  {Choudhury}} \emph {et~al.},\ }\href@noop {} {}\bibinfo {note} {Technical
  Design Report of the Forward Silicon Vertex Tracker, 2007}\BibitemShut
  {NoStop}%
\bibitem [{\citenamefont {Olive}\ \emph {et~al.}(2014)\citenamefont {Olive}
  \emph {et~al.}}]{Agashe:2014kda}%
  \BibitemOpen
  \bibfield  {author} {\bibinfo {author} {\bibfnamefont {K.~A.}\ \bibnamefont
  {Olive}} \emph {et~al.} (\bibinfo {collaboration} {Particle Data Group}),\
  }\bibfield  {title} {\enquote {\bibinfo {title} {{Rev. of Particle Phys.}}}\
  }\href {\doibase 10.1088/1674-1137/38/9/090001} {\bibfield  {journal}
  {\bibinfo  {journal} {Chin. Phys. C}\ }\textbf {\bibinfo {volume} {38}},\
  \bibinfo {pages} {090001} (\bibinfo {year} {2014})}\BibitemShut {NoStop}%
\bibitem [{\citenamefont {Agostinelli}\ \emph {et~al.}(2003)\citenamefont
  {Agostinelli} \emph {et~al.}}]{Agostinelli:2002hh}%
  \BibitemOpen
  \bibfield  {author} {\bibinfo {author} {\bibfnamefont {S.}~\bibnamefont
  {Agostinelli}} \emph {et~al.} (\bibinfo {collaboration} {GEANT4
  Collaboration}),\ }\bibfield  {title} {\enquote {\bibinfo {title} {{GEANT4: A
  Simulation toolkit}},}\ }\href {\doibase 10.1016/S0168-9002(03)01368-8}
  {\bibfield  {journal} {\bibinfo  {journal} {Nucl. Instrum. Methods Phys.
  Res., Sec. A}\ }\textbf {\bibinfo {volume} {506}},\ \bibinfo {pages}
  {250--303} (\bibinfo {year} {2003})}\BibitemShut {NoStop}%
\bibitem [{\citenamefont {Adare}\ \emph {et~al.}(2011)\citenamefont {Adare}
  \emph {et~al.}}]{Adare:2010fe}%
  \BibitemOpen
  \bibfield  {author} {\bibinfo {author} {\bibfnamefont {A.}~\bibnamefont
  {Adare}} \emph {et~al.} (\bibinfo {collaboration} {PHENIX Collaboration}),\
  }\bibfield  {title} {\enquote {\bibinfo {title} {{Measurement of neutral
  mesons in p+p collisions at $\sqrt(s)$= 200 GeV and scaling properties of
  hadron production}},}\ }\href {\doibase 10.1103/PhysRevD.83.052004}
  {\bibfield  {journal} {\bibinfo  {journal} {Phys. Rev. D}\ }\textbf {\bibinfo
  {volume} {83}},\ \bibinfo {pages} {052004} (\bibinfo {year}
  {2011})}\BibitemShut {NoStop}%
\bibitem [{\citenamefont {Adare}\ \emph {et~al.}(2010)\citenamefont {Adare}
  \emph {et~al.}}]{Adare:2009js}%
  \BibitemOpen
  \bibfield  {author} {\bibinfo {author} {\bibfnamefont {A.}~\bibnamefont
  {Adare}} \emph {et~al.} (\bibinfo {collaboration} {PHENIX Collaboration}),\
  }\bibfield  {title} {\enquote {\bibinfo {title} {{Transverse momentum
  dependence of $J/\psi$ polarization at midrapidity in $p+p$ collisions at
  $\sqrt{s}=200$ GeV}},}\ }\href {\doibase 10.1103/PhysRevD.82.012001}
  {\bibfield  {journal} {\bibinfo  {journal} {Phys. Rev. D}\ }\textbf {\bibinfo
  {volume} {82}},\ \bibinfo {pages} {012001} (\bibinfo {year}
  {2010})}\BibitemShut {NoStop}%
\bibitem [{\citenamefont {Antoniazzi}\ \emph {et~al.}(1993)\citenamefont
  {Antoniazzi} \emph {et~al.}}]{Antoniazzi:1992iv}%
  \BibitemOpen
  \bibfield  {author} {\bibinfo {author} {\bibfnamefont {L.}~\bibnamefont
  {Antoniazzi}} \emph {et~al.} (\bibinfo {collaboration} {E705
  Collaboration}),\ }\bibfield  {title} {\enquote {\bibinfo {title}
  {{Production of $J/\psi$ via $\psi^\prime$ and $\chi$ decay in 300 GeV/c $p$
  and $\pi^{\pm}$ nucleon interactions}},}\ }\href {\doibase
  10.1103/PhysRevLett.70.383} {\bibfield  {journal} {\bibinfo  {journal} {Phys.
  Rev. Lett.}\ }\textbf {\bibinfo {volume} {70}},\ \bibinfo {pages} {383--386}
  (\bibinfo {year} {1993})}\BibitemShut {NoStop}%
\bibitem [{\citenamefont {Abreu}\ \emph {et~al.}(1998)\citenamefont {Abreu}
  \emph {et~al.}}]{Abreu:1998rx}%
  \BibitemOpen
  \bibfield  {author} {\bibinfo {author} {\bibfnamefont {M.~C.}\ \bibnamefont
  {Abreu}} \emph {et~al.} (\bibinfo {collaboration} {NA51 Collaboration}),\
  }\bibfield  {title} {\enquote {\bibinfo {title} {{$J/\psi$, $\psi^\prime$ and
  Drell-Yan production in $p+p$ and $p+d$ interactions at 450 GeV/$c$}},}\
  }\href {\doibase 10.1016/S0370-2693(98)01014-4} {\bibfield  {journal}
  {\bibinfo  {journal} {Phys. Lett. B}\ }\textbf {\bibinfo {volume} {438}},\
  \bibinfo {pages} {35--40} (\bibinfo {year} {1998})}\BibitemShut {NoStop}%
\bibitem [{\citenamefont {Clark}\ \emph {et~al.}(1978)\citenamefont {Clark}
  \emph {et~al.}}]{Clark:1978mg}%
  \BibitemOpen
  \bibfield  {author} {\bibinfo {author} {\bibfnamefont {A.~G.}\ \bibnamefont
  {Clark}} \emph {et~al.},\ }\bibfield  {title} {\enquote {\bibinfo {title}
  {{Electron Pair Production at the CERN ISR}},}\ }\href {\doibase
  10.1016/0550-3213(78)90400-5} {\bibfield  {journal} {\bibinfo  {journal}
  {Nucl. Phys. B}\ }\textbf {\bibinfo {volume} {142}},\ \bibinfo {pages} {29}
  (\bibinfo {year} {1978})}\BibitemShut {NoStop}%
\bibitem [{\citenamefont {Albajar}\ \emph {et~al.}(1991)\citenamefont {Albajar}
  \emph {et~al.}}]{Albajar:1990hf}%
  \BibitemOpen
  \bibfield  {author} {\bibinfo {author} {\bibfnamefont {C.}~\bibnamefont
  {Albajar}} \emph {et~al.} (\bibinfo {collaboration} {UA1 Collaboration}),\
  }\bibfield  {title} {\enquote {\bibinfo {title} {{$J/\psi$ and$\psi^\prime$
  production at the CERN $p\bar{p}$ collider}},}\ }\href {\doibase
  10.1016/0370-2693(91)90227-H} {\bibfield  {journal} {\bibinfo  {journal}
  {Phys. Lett. B}\ }\textbf {\bibinfo {volume} {256}},\ \bibinfo {pages}
  {112--120} (\bibinfo {year} {1991})}\BibitemShut {NoStop}%
\bibitem [{\citenamefont {Abe}\ \emph {et~al.}(1997)\citenamefont {Abe} \emph
  {et~al.}}]{Abe:1997jz}%
  \BibitemOpen
  \bibfield  {author} {\bibinfo {author} {\bibfnamefont {F.}~\bibnamefont
  {Abe}} \emph {et~al.} (\bibinfo {collaboration} {CDF Collaboration}),\
  }\bibfield  {title} {\enquote {\bibinfo {title} {{$J/\psi$ and $\psi(2S)$
  production in $p\bar{p}$ collisions at $\sqrt{s}=1.8$ TeV}},}\ }\href
  {\doibase 10.1103/PhysRevLett.79.572} {\bibfield  {journal} {\bibinfo
  {journal} {Phys. Rev. Lett.}\ }\textbf {\bibinfo {volume} {79}},\ \bibinfo
  {pages} {572--577} (\bibinfo {year} {1997})}\BibitemShut {NoStop}%
\bibitem [{\citenamefont {Aaij}\ \emph {et~al.}(2013)\citenamefont {Aaij} \emph
  {et~al.}}]{Aaij:2013jxj}%
  \BibitemOpen
  \bibfield  {author} {\bibinfo {author} {\bibfnamefont {R}~\bibnamefont
  {Aaij}} \emph {et~al.} (\bibinfo {collaboration} {LHCb Collaboration}),\
  }\bibfield  {title} {\enquote {\bibinfo {title} {{Exclusive $J/\psi$ and
  $\psi$(2S) production in $pp$ collisions at $ \sqrt{s}=7$ TeV}},}\ }\href
  {\doibase 10.1088/0954-3899/40/4/045001} {\bibfield  {journal} {\bibinfo
  {journal} {J. Phys. G}\ }\textbf {\bibinfo {volume} {40}},\ \bibinfo {pages}
  {045001} (\bibinfo {year} {2013})}\BibitemShut {NoStop}%
\bibitem [{\citenamefont {Abelev}\ \emph {et~al.}(2014)\citenamefont {Abelev}
  \emph {et~al.}}]{Abelev:2014qha}%
  \BibitemOpen
  \bibfield  {author} {\bibinfo {author} {\bibfnamefont {B.~B.}\ \bibnamefont
  {Abelev}} \emph {et~al.} (\bibinfo {collaboration} {ALICE Collaboration}),\
  }\bibfield  {title} {\enquote {\bibinfo {title} {{Measurement of quarkonium
  production at forward rapidity in $pp$ collisions at $\sqrt{s}=7$ TeV}},}\
  }\href {\doibase 10.1140/epjc/s10052-014-2974-4} {\bibfield  {journal}
  {\bibinfo  {journal} {Eur. Phys. J. C}\ }\textbf {\bibinfo {volume} {74}},\
  \bibinfo {pages} {2974} (\bibinfo {year} {2014})}\BibitemShut {NoStop}%
\bibitem [{\citenamefont {Aaij}\ \emph {et~al.}(2012)\citenamefont {Aaij} \emph
  {et~al.}}]{Aaij:2012ag}%
  \BibitemOpen
  \bibfield  {author} {\bibinfo {author} {\bibfnamefont {R}~\bibnamefont
  {Aaij}} \emph {et~al.} (\bibinfo {collaboration} {LHCb Collaboration}),\
  }\bibfield  {title} {\enquote {\bibinfo {title} {{Measurement of $\psi(2S)$
  meson production in $pp$ collisions at $\sqrt{s}=7$ TeV}},}\ }\href {\doibase
  10.1140/epjc/s10052-012-2100-4} {\bibfield  {journal} {\bibinfo  {journal}
  {Eur. Phys. J. C}\ }\textbf {\bibinfo {volume} {72}},\ \bibinfo {pages}
  {2100} (\bibinfo {year} {2012})},\ \Eprint {http://arxiv.org/abs/1204.1258}
  {1204.1258} \BibitemShut {NoStop}%
\bibitem [{\citenamefont {Abt}\ \emph {et~al.}(2007)\citenamefont {Abt} \emph
  {et~al.}}]{Abt:2006va}%
  \BibitemOpen
  \bibfield  {author} {\bibinfo {author} {\bibfnamefont {I.}~\bibnamefont
  {Abt}} \emph {et~al.} (\bibinfo {collaboration} {HERA-B Collaboration}),\
  }\bibfield  {title} {\enquote {\bibinfo {title} {{A Measurement of the
  $\psi^\prime$ to $J/\psi$ production ratio in 920 GeV proton-nucleus
  interactions}},}\ }\href {\doibase 10.1140/epjc/s10052-006-0139-9} {\bibfield
   {journal} {\bibinfo  {journal} {Eur. Phys. J. C}\ }\textbf {\bibinfo
  {volume} {49}},\ \bibinfo {pages} {545--558} (\bibinfo {year}
  {2007})}\BibitemShut {NoStop}%
\bibitem [{\citenamefont {Schub}\ \emph {et~al.}(1995)\citenamefont {Schub}
  \emph {et~al.}}]{Schub:1995pu}%
  \BibitemOpen
  \bibfield  {author} {\bibinfo {author} {\bibfnamefont {M.~H.}\ \bibnamefont
  {Schub}} \emph {et~al.} (\bibinfo {collaboration} {E789 Collaboration}),\
  }\bibfield  {title} {\enquote {\bibinfo {title} {{Measurement of $J/\psi$ and
  $\psi^\prime$ production in 800 GeV/c proton-gold collisions}},}\ }\href
  {\doibase 10.1103/PhysRevD.53.570, 10.1103/PhysRevD.52.1307} {\bibfield
  {journal} {\bibinfo  {journal} {Phys. Rev. D}\ }\textbf {\bibinfo {volume}
  {52}},\ \bibinfo {pages} {1307} (\bibinfo {year} {1995})},\ \bibinfo {note}
  {[Erratum: Phys. Rev.D53,570(1996)]}\BibitemShut {NoStop}%
\bibitem [{\citenamefont {Alver}\ \emph {et~al.}(2011)\citenamefont {Alver}
  \emph {et~al.}}]{Alver:2010ck}%
  \BibitemOpen
  \bibfield  {author} {\bibinfo {author} {\bibfnamefont {B.}~\bibnamefont
  {Alver}} \emph {et~al.} (\bibinfo {collaboration} {PHOBOS Collaboration}),\
  }\bibfield  {title} {\enquote {\bibinfo {title} {{Phobos results on charged
  particle multiplicity and pseudorapidity distributions in Au+Au, Cu+Cu,
  $d$+Au, and $p+p$ collisions at ultra-relativistic energies}},}\ }\href
  {\doibase 10.1103/PhysRevC.83.024913} {\bibfield  {journal} {\bibinfo
  {journal} {Phys. Rev. C}\ }\textbf {\bibinfo {volume} {83}},\ \bibinfo
  {pages} {024913} (\bibinfo {year} {2011})}\BibitemShut {NoStop}%
\bibitem [{\citenamefont {Ferreiro}(2016)}]{Ferreiro_privcomm}%
  \BibitemOpen
  \bibfield  {author} {\bibinfo {author} {\bibfnamefont {E.~G.}\ \bibnamefont
  {Ferreiro}},\ }\href@noop {} {} (\bibinfo {year} {2016}),\ \bibinfo {note}
  {private communication}\BibitemShut {NoStop}%
\bibitem [{\citenamefont {Abelev}\ \emph {et~al.}()\citenamefont {Abelev} \emph
  {et~al.}}]{Abelev:2014zpa}%
  \BibitemOpen
  \bibfield  {author} {\bibinfo {author} {\bibfnamefont {B.~B.}\ \bibnamefont
  {Abelev}} \emph {et~al.} (\bibinfo {collaboration} {ALICE Collaboration}),\
  }\href@noop {} {\enquote {\bibinfo {title} {{Suppression of $\psi$(2S)
  production in $p$-Pb collisions at $\sqrt{s_{NN}}=5.02$ TeV}},}\ }\bibinfo
  {note} {{J. High Energy Phys. {\bf 12 (2014)} 073}}\BibitemShut {NoStop}%
\bibitem [{\citenamefont {Aaij}\ \emph {et~al.}()\citenamefont {Aaij} \emph
  {et~al.}}]{Aaij:2016eyl}%
  \BibitemOpen
  \bibfield  {author} {\bibinfo {author} {\bibfnamefont {R.}~\bibnamefont
  {Aaij}} \emph {et~al.} (\bibinfo {collaboration} {LHCb Collaboration}),\
  }\href@noop {} {\enquote {\bibinfo {title} {{Study of $\psi(2S)$ production
  and cold nuclear matter effects in pPb collisions at $\sqrt{s_{NN}}=5$
  TeV}},}\ }\bibinfo {note} {{J. High Energy Phys. {\bf 03 (2016)}
  133}}\BibitemShut {NoStop}%
\bibitem [{\citenamefont {Abelev}\ \emph {et~al.}(2013)\citenamefont {Abelev}
  \emph {et~al.}}]{ALICE:2012xs}%
  \BibitemOpen
  \bibfield  {author} {\bibinfo {author} {\bibfnamefont {B.}~\bibnamefont
  {Abelev}} \emph {et~al.} (\bibinfo {collaboration} {ALICE Collaboration}),\
  }\bibfield  {title} {\enquote {\bibinfo {title} {{Pseudorapidity density of
  charged particles in $p$+Pb collisions at $\sqrt{s_{NN}}=5.02$ TeV}},}\
  }\href {\doibase 10.1103/PhysRevLett.110.032301} {\bibfield  {journal}
  {\bibinfo  {journal} {Phys. Rev. Lett.}\ }\textbf {\bibinfo {volume} {110}},\
  \bibinfo {pages} {032301} (\bibinfo {year} {2013})}\BibitemShut {NoStop}%
\bibitem [{\citenamefont {Lin}\ \emph {et~al.}(2002)\citenamefont {Lin},
  \citenamefont {Ko},\ and\ \citenamefont {Pal}}]{Lin:2002gc}%
  \BibitemOpen
  \bibfield  {author} {\bibinfo {author} {\bibfnamefont {Z.~W.}\ \bibnamefont
  {Lin}}, \bibinfo {author} {\bibfnamefont {C.~M.}\ \bibnamefont {Ko}}, \ and\
  \bibinfo {author} {\bibfnamefont {S.}~\bibnamefont {Pal}},\ }\bibfield
  {title} {\enquote {\bibinfo {title} {{Partonic effects on pion interferometry
  at RHIC}},}\ }\href {\doibase 10.1103/PhysRevLett.89.152301} {\bibfield
  {journal} {\bibinfo  {journal} {Phys. Rev. Lett.}\ }\textbf {\bibinfo
  {volume} {89}},\ \bibinfo {pages} {152301} (\bibinfo {year}
  {2002})}\BibitemShut {NoStop}%
\bibitem [{\citenamefont {Lin}\ \emph {et~al.}(2005)\citenamefont {Lin}, ,
  \citenamefont {Ko}, \citenamefont {Li}, \citenamefont {Zhang},\ and\
  \citenamefont {Pal}}]{Lin:2004en}%
  \BibitemOpen
  \bibfield  {author} {\bibinfo {author} {\bibfnamefont {Z.~W.}\ \bibnamefont
  {Lin}}, , \bibinfo {author} {\bibfnamefont {C.~M.}\ \bibnamefont {Ko}},
  \bibinfo {author} {\bibfnamefont {B.~A.}\ \bibnamefont {Li}}, \bibinfo
  {author} {\bibfnamefont {B.}~\bibnamefont {Zhang}}, \ and\ \bibinfo {author}
  {\bibfnamefont {S.}~\bibnamefont {Pal}},\ }\bibfield  {title} {\enquote
  {\bibinfo {title} {{A Multi-phase transport model for relativistic heavy ion
  collisions}},}\ }\href {\doibase 10.1103/PhysRevC.72.064901} {\bibfield
  {journal} {\bibinfo  {journal} {Phys. Rev. C}\ }\textbf {\bibinfo {volume}
  {72}},\ \bibinfo {pages} {064901} (\bibinfo {year} {2005})}\BibitemShut
  {NoStop}%
\bibitem [{\citenamefont {Loizides}\ \emph {et~al.}(2015)\citenamefont
  {Loizides}, \citenamefont {Nagle},\ and\ \citenamefont
  {Steinberg}}]{Loizides:2014vua}%
  \BibitemOpen
  \bibfield  {author} {\bibinfo {author} {\bibfnamefont {C.}~\bibnamefont
  {Loizides}}, \bibinfo {author} {\bibfnamefont {J.}~\bibnamefont {Nagle}}, \
  and\ \bibinfo {author} {\bibfnamefont {P.}~\bibnamefont {Steinberg}},\
  }\bibfield  {title} {\enquote {\bibinfo {title} {{Improved version of the
  PHOBOS Glauber Monte Carlo}},}\ }\href {\doibase 10.1016/j.softx.2015.05.001}
  {\bibfield  {journal} {\bibinfo  {journal} {SoftwareX}\ }\textbf {\bibinfo
  {volume} {1-2}},\ \bibinfo {pages} {13--18} (\bibinfo {year}
  {2015})}\BibitemShut {NoStop}%
\bibitem [{\citenamefont {Nagle}\ \emph {et~al.}(2014)\citenamefont {Nagle},
  \citenamefont {Adare}, \citenamefont {Beckman}, \citenamefont {Koblesky},
  \citenamefont {Koop}, \citenamefont {McGlinchey}, \citenamefont {Romatschke},
  \citenamefont {Carlson}, \citenamefont {Lynn},\ and\ \citenamefont
  {McCumber}}]{Nagle:2013lja}%
  \BibitemOpen
  \bibfield  {author} {\bibinfo {author} {\bibfnamefont {J.~L.}\ \bibnamefont
  {Nagle}}, \bibinfo {author} {\bibfnamefont {A.}~\bibnamefont {Adare}},
  \bibinfo {author} {\bibfnamefont {S.}~\bibnamefont {Beckman}}, \bibinfo
  {author} {\bibfnamefont {T.}~\bibnamefont {Koblesky}}, \bibinfo {author}
  {\bibfnamefont {J.~O.}\ \bibnamefont {Koop}}, \bibinfo {author}
  {\bibfnamefont {D.}~\bibnamefont {McGlinchey}}, \bibinfo {author}
  {\bibfnamefont {P.}~\bibnamefont {Romatschke}}, \bibinfo {author}
  {\bibfnamefont {J.}~\bibnamefont {Carlson}}, \bibinfo {author} {\bibfnamefont
  {J.~E.}\ \bibnamefont {Lynn}}, \ and\ \bibinfo {author} {\bibfnamefont
  {M.}~\bibnamefont {McCumber}},\ }\bibfield  {title} {\enquote {\bibinfo
  {title} {{Exploiting Intrinsic Triangular Geometry in Relativistic
  $^{3}$He+Au Collisions to Disentangle Medium Properties}},}\ }\href {\doibase
  10.1103/PhysRevLett.113.112301} {\bibfield  {journal} {\bibinfo  {journal}
  {Phys. Rev. Lett.}\ }\textbf {\bibinfo {volume} {113}},\ \bibinfo {pages}
  {112301} (\bibinfo {year} {2014})}\BibitemShut {NoStop}%
\bibitem [{\citenamefont {Adam}\ \emph {et~al.}()\citenamefont {Adam} \emph
  {et~al.}}]{Adam:2015isa}%
  \BibitemOpen
  \bibfield  {author} {\bibinfo {author} {\bibfnamefont {J.}~\bibnamefont
  {Adam}} \emph {et~al.} (\bibinfo {collaboration} {ALICE Collaboration}),\
  }\href@noop {} {\enquote {\bibinfo {title} {{Differential studies of
  inclusive $J/\psi$ and $\psi(2S)$ production at forward rapidity in Pb-Pb
  collisions at $\sqrt{s_{NN}}=2.76$ TeV}},}\ }\bibinfo {note} {{J. High Energy
  Phys. {\bf 05 (2016)} 179}}\BibitemShut {NoStop}%
\bibitem [{\citenamefont {Khachatryan}\ \emph {et~al.}(2014)\citenamefont
  {Khachatryan} \emph {et~al.}}]{Khachatryan:2014bva}%
  \BibitemOpen
  \bibfield  {author} {\bibinfo {author} {\bibfnamefont {V.}~\bibnamefont
  {Khachatryan}} \emph {et~al.} (\bibinfo {collaboration} {CMS
  Collaboration}),\ }\bibfield  {title} {\enquote {\bibinfo {title}
  {{Measurement of Prompt $\psi(2S) \to J/\psi$ Yield Ratios in Pb-Pb and $p-p$
  Collisions at $\sqrt {s_{NN}}=2.76$ TeV}},}\ }\href {\doibase
  10.1103/PhysRevLett.113.262301} {\bibfield  {journal} {\bibinfo  {journal}
  {Phys. Rev. Lett.}\ }\textbf {\bibinfo {volume} {113}},\ \bibinfo {pages}
  {262301} (\bibinfo {year} {2014})}\BibitemShut {NoStop}%
\bibitem [{\citenamefont {Chatrchyan}\ \emph {et~al.}(2012)\citenamefont
  {Chatrchyan} \emph {et~al.}}]{Chatrchyan:2012lxa}%
  \BibitemOpen
  \bibfield  {author} {\bibinfo {author} {\bibfnamefont {S.}~\bibnamefont
  {Chatrchyan}} \emph {et~al.} (\bibinfo {collaboration} {CMS Collaboration}),\
  }\bibfield  {title} {\enquote {\bibinfo {title} {{Observation of sequential
  Upsilon suppression in PbPb collisions}},}\ }\href {\doibase
  10.1103/PhysRevLett.109.222301} {\bibfield  {journal} {\bibinfo  {journal}
  {Phys. Rev. Lett.}\ }\textbf {\bibinfo {volume} {109}},\ \bibinfo {pages}
  {222301} (\bibinfo {year} {2012})}\BibitemShut {NoStop}%
\end{thebibliography}

%
 
\end{document}